\documentclass[12pt]{article}
\usepackage{graphicx}
\usepackage{amssymb}
\def\beq{\begin{equation}}
\def\eeq{\end{equation}}
\def\bea{\begin{eqnarray}}
\def\eea{\end{eqnarray}}

\def\roughly#1{\mathrel{\raise.3ex\hbox
{$#1$\kern-.75em\lower1ex\hbox{$\sim$}}}}

          \def\s{\tau}
     \def\t{\tilde}        
  \def\rr2{{1\over\sqrt{2}}}

\def\.{\!\cdot\!}    \def\:{\cdots}   \def\[{\left[}   \def\]{\right]}
\def\({\left(} \def\){\right)} 
\def\t3{{1\over\sqrt{3}}}
\def\s6{{1\over\sqrt{6}}}
\def\t13{ \theta_{13}}

\begin{document}
\begin{flushright}
UMISS-HEP-2011-04 \\
[10mm]
\end{flushright}

\begin{center}
\bigskip {\Large  \bf  The charged lepton mass matrix and non-zero 
$\theta_{13}$ with TeV scale New Physics.}  
\\[8mm]
Ahmed Rashed $^{\dag, \ddag}$
\footnote{E-mail:
\texttt{amrashed@phy.olemiss.edu}}
and Alakabha Datta $^{\dag}$ 
\footnote{E-mail:
\texttt{datta@phy.olemiss.edu}} 
\end{center}

\begin{center}
~~~~ {\it $^{\dag}$ Department  of Physics and Astronomy,}\\ 
~~~~~{ \it University of Mississippi,}\\
~~~~~{\it  Lewis Hall, University, MS, 38677.}\\
\end{center}

\begin{center}
~~~~ {\it $^{\ddag}$ Department  of Physics,}\\ 
~~~~~{ \it Faculty of Science,}\\
~~~~~{\it  Ain Shams University, Cairo, 11566, Egypt.}\\
                             \end{center}

\begin{center} 
\bigskip (\today) \vskip0.5cm {\Large Abstract\\} \vskip3truemm
\parbox[t]{\textwidth} {
 We provide an explicit structure of the charged lepton mass matrix which is 2-3 symmetric except for a single breaking of this symmetry by the muon mass. We identify a flavor symmetric limit for the mass matrices where the first generation is decoupled from the other two in the charged lepton sector while in the neutrino sector the third generation is decoupled from the first two generations. The leptonic mixing in the symmetric limit can be, among other structures, the bi-maximal (BM) or the tri-bimaximal (TBM) mixing. Symmetry breaking effects are included both in the charged lepton and the neutrino sector to produce corrections to the leptonic mixing and explain the recent $\theta_{13}$ measurements. A model that extends the SM by three right handed neutrinos, an extra Higgs doublet, and two singlet scalars is introduced to generate the leptonic mixing.
}
\end{center}

\thispagestyle{empty} \newpage \setcounter{page}{1}
\baselineskip=14pt


\section{Introduction}


We now know that neutrinos have masses and just like the quark mixing matrix
there is a leptonic mixing matrix. This fact has been firmly established through a
variety of solar, atmospheric, and terrestrial neutrino oscillation experiments \cite{Bahcall:2004cc-1}. We  parametrize the neutrino mixing matrix, $U_{PMNS}$, as follows \cite{PMNS-1}:
\begin{eqnarray}
 U_{PMNS}=\pmatrix{c_{12}c_{13} & s_{12}c_{13} &
s_{13} e^{-i\delta} \cr
-s_{12}c_{23}-c_{12}s_{23}s_{13}e^{i\delta}
&c_{12}c_{23}-s_{12}s_{23}s_{13}e^{i\delta} & s_{23}c_{13}\cr
s_{12}s_{23}-c_{12}c_{23}s_{13}e^{i\delta}
&-c_{12}s_{23}-s_{12}c_{23}s_{13}e^{i\delta} &
 c_{23}c_{13}}K,
 \label{PMNS2}
\end{eqnarray} 
where $s_{13}\equiv\sin \theta_{13}$, $c_{13}\equiv\cos \theta_{13}$
with $\theta_{13}$ being the
reactor angle, $s_{12}\equiv\sin \theta_{12}$, $c_{12}\equiv\cos \theta_{12}$
with $\theta_{12}$ being the
solar angle, $s_{23}\equiv\sin \theta_{23}$, $c_{23}\equiv\cos \theta_{23}$
with $\theta_{23}$ being the
atmospheric angle, $\delta$ is the Dirac CP violating phase, and
$K~=~diag(1, e^{i\phi_1},e^{i\phi_2})$
 contains additional (Majorana)
CP violating phases $\phi_1, \phi_2$. We  ignore the
Majorana CP violating phases in this work.

Unlike the CKM matrix that can be thought of as a perturbation about the identity matrix the leading 
term in the leptonic mixing contains large mixing angles. Some examples of the leading order mixing matrix are the bi-maximal mixing \cite{BM-1} and the tri-bimaximal mixing \cite{TBM-1}. However, current
 experiments indicate deviations from these standard zeroth order forms.

{} For instance, recent results from the T2K \cite{Abe:2011sj} and MINOS \cite{MINOS, MINOS-formal} experiments 
have indicated a large reactor angle $\t13$ for neutrino mixing. At the 90\% C.L., T2K gives 
$0.03 (0.04) < \sin^2 2 \theta_{13} < 0.28 (0.34)$, with zero Dirac {\tt CP} phase, $\delta_D$, 
for normal (inverted) hierarchy. The MINOS group gives $0.01(0.026) < \sin^2 2 \theta_{13} < 0.088(0.150)$. 
There are already several papers that have attempted to explain the resent $\t13$ results \cite{T2K-cited-CL-NonDiag-1, T2K-cited-CL-Diag-2}. Large $\theta_{13}$ was anticipated in Ref.~\cite{He:2011kn}.

The leptonic mixing arises from the overlap of matrices that diagonalize the charged lepton and the neutrino mass matrices. Many approaches to studying the leptonic mixing start in the basis where the
charged lepton mass is diagonal.
Our approach to obtaining the leading order leptonic mixing as well as deviations from  it starts from the charged lepton sector. 
A recent attempt to understand $\t13$ from the charged lepton sector can be found
in Ref.~\cite{T2K-cited-CL-NonDiag-1} and in the past corrections to the leptonic mixing from the charged lepton sector were considered in Ref.~\cite{CCC-1}.
An approach to suppress flavor changing neutral current effects (FCNC) in the quark sector, based on shared flavor symmetry, was proposed in Ref.~\cite{datta-1}. As an example of this shared symmetry  the decoupled $2-3$ symmetry  was used for the down quark sector to suppress FCNC effects
and explain anomalies \cite{datta1-1} observed in the $B$ meson system. In the decoupled limit the first generation is decoupled from the other two generations. We extend this decoupled $2-3$ symmetry to the charged lepton sector. This is a reasonable extension given  the fact that the down quark and charged leptons exhibit similar hierarchical structure and they may be combined in representations of GUT groups.
 
 One of the central ideas of this approach is the requirement that the mass matrices, in a symmetric limit,
 be diagonalized by unitary matrices composed of pure numbers  independent of the
parameters of the mass matrices. This is similar to the idea of form diagonalizable matrices discussed in Ref.~\cite{TBM-5}.
 If one starts with a $2-3$ symmetric mass matrix for the charged lepton sector and requires it to be diagonalized by unitary matrices of pure numbers one recovers the decoupled $2-3$ symmetry.
 In the neutrino sector we assume the third generation to be decoupled from the first two generations. With real entries in the neutrino mass matrix it is diagonalized by a rotation matrix and the resulting leptonic mixing has a $\mu-\tau$ symmetry. Requiring the mass matrix to be diagonalized by pure numbers can lead to, among other structures, the BM and the TBM leptonic mixing.
 
To generate the mixing matrices in the charged lepton and the neutrino sector, we present a Lagrangian that extends the SM by three right handed neutrinos, an additional Higgs doublet and two singlet scalar fields.{\footnote{ Recent motivations for considering two Higgs doublet models can be found in 
Ref.~\cite{2HDM-1}.}} The Lagrangian uses the same class of $Z_2$ symmetries as has been used in  Ref~\cite{Grimus:2003kq-1}.
However, the structure as well as the phenomenology of our model is very different
from the above mentioned papers.
 The Lagrangian is constructed to have  a $2-3$ symmetry, $Z_{2}^{23}$, along with two additional $Z_2$ symmetries
$Z_2^e$ and $Z_{2}^D$ . The neutrino masses and mixing are generated through the usual see-saw mechanism. The presence of the  $Z_2^{23} \times Z_2^e$ symmetries lead to the decoupled $2-3$ symmetry in the charged lepton sector and fixes the interactions of the right handed neutrinos with the singlet scalar fields. The presence of the $Z_2^D$ symmetry
 forces the  neutrinos to acquire Dirac masses by coupling to a second Higgs doublet which has a different $Z_2^D$ transformation  than the usual SM Higgs doublet that give masses to the charged leptons.  The full Lagrangian is symmetric under the product of the $Z_2$ symmetries,
 $Z_2^{23} \times Z_2^e \times Z_2^D$.

 The neutrino masses and mixing arise when the Higgs doublets and the singlet scalars acquire v.e.v's and break the symmetries of the Lagrangian.
 The leptonic mixing is predicted to be of the bi-maximal type  when both the singlet scalars acquire the same v.e.v. If the v.e.v of the second Higgs doublet is small enough $ \sim MeV$ then the see-saw scale as well as the masses of the singlet scalars can be in the TeV range.
To obtain the TBM mixing one has to use different flavor symmetries.

Symmetry breaking is introduced in the charged lepton sector  by higher dimensional operators that break the decoupled $2-3$ symmetry but generate a $2-3$ symmetric mass matrix except for a 
single breaking generated by the muon mass. In the neutrino sector, symmetry breaking is introduced
by breaking the alignment of the v.e.v's of the singlet scalars by terms in the effective potential.
The corrections to leptonic mixing go as $\sim {v^2 \over \omega^2}$ where $v$ is the v.e.v of the SM Higgs and $\omega$ the scale of the singlet scalar v.e.v's. If $ \omega \sim $ TeV then the corrections to the leptonic mixing are enough to explain the experimental observations.

The paper is organized in the following manner: We begin in Sec.~2 with a
discussion of the flavor symmetric limit that leads to among other structures the BM and TBM mixing.
 In Sec.~3 we present the Lagrangian to generate the mixing matrices in the symmetric limit. In Sec.~4 we study the effect of symmetry breaking in the charged lepton and neutrino
sector to generate the realistic leptonic mixing matrix. In Sec.~5 we show the
numerical results due to the symmetry breaking, and, finally, in Sec.~6 we conclude
with a summary of the results reported in this work.


\section{The Leptonic Mixing in the Symmetric Limit}


We start with the charged lepton sector, and assume that the Yukawa matrix is $2-3$ symmetric \cite{mutau-1}. The Yukawa couplings of the charged leptons are given by
\bea
Y^L & = &\pmatrix{l_{11} & l_{12}
&  -l_{12}  \cr  l_{12}  &  l_{22}  &  l_{23}\cr  -l_{12}  &l_{23}  &l_{22}
}. \
\label{PMNS}
\eea
The above Yukawa matrix can be diagonalized as
\bea
U^{\dagger} Y^L U & = & Y^L_{diag}, \nonumber\\
U & = &\pmatrix{1 & 0
&  0  \cr  0  & \frac{1}{\sqrt{2}}   &  \frac{1}{\sqrt{2}}\cr  0  &-\frac{1}{\sqrt{2}}  &\frac{1}{\sqrt{2}}
} 
\cdot
\pmatrix{\cos{\theta} & \sin{\theta}
&  0  \cr  -\sin{\theta}  & \cos{\theta}   &  0\cr  0  &0  &1
},\ 
\label{lam}
\eea
where the mixing angle $\theta$ is determined by the positive solution to
 \bea
 \tan \theta  & = & {2 \sqrt{2}l_{12} \over {
 l_{22}-l_{23}-l_{11} \pm 
 \sqrt{ (l_{22}-l_{23}-l_{11})^2+8l_{12}^2}
 }
 }. \
\label{Tan}
 \eea
 The eigenvalues of $Y^L$ are ${ 1 \over 2} [l_{11}+l_{22}-l_{23} \pm \sqrt{(l_{11}-l_{22}+l_{23})^2+8l_{12}^2} ]$ and $l_{22}+l_{23}$. According to our assumption, the elements of the matrix that diagonalizes $Y^L$ must be pure numbers in the symmetric limit. It is clear that we can achieve that
 by setting $l_{12}=0$ ($\theta =0$) in Eq.~\ref{Tan}. This generates  the decoupled 2-3 symmetry \cite{datta-1}, as the flavor symmetry in the charged lepton sector in which the
 first generation is decoupled from the second and third generations. 
 
  One can represent the Yukawa matrix with the decoupled 2-3 symmetry by $Y^L_{23}$ as 
\bea
Y^L_{23} & = &
\pmatrix{l_{11} & 0
&  0  \cr  0  &  \frac{1}{2}{l_{22}}  &  \frac{1}{2}{l_{23}}\cr  0  
&\frac{1}{2}{l_{23}}  &\frac{1}{2}{l_{22}}}. \
 \label{23sym}
 \eea
This Yukawa matrix  $Y^L_{23}$ is diagonalized by the unitary matrix $W^{l}_{23}$ given by
\bea
W^l_{23} &  = & \pmatrix{1 & 0
&   0  \cr   0   &  -\frac{1}{\sqrt{2}}   &  \frac{1}{\sqrt{2}}\cr   0
&\frac{1}{\sqrt{2}} &  \frac{1}{\sqrt{2}}}.\
\label{wl} 
\eea 
Note that this matrix differs from the one in Eq.~\ref{lam} in the limit $\theta=0$ by an irrelevant diagonal phase matrix. 
Writing the diagonalized Yukawa matrix as $Y^{L}_{23d}$  we have
\bea 
Y^L_{23d} & = & W^{l \dagger}_{23} Y^L_{23} W^l_{23} 
=\pmatrix{l_{11} & 0
&  0  \cr  0  &   \frac{1}{2}{(l_{22}-l_{23})}  &  0\cr  0  &0  &
\frac{1}{2}{(l_{22}+l_{23})}}. \  
\label{diag23sym} 
\eea
The  charged lepton  masses   are  given   by
  \bea  m_e   &  =   &  \pm
\frac{v_1}{\sqrt{2}}l_{11},       \nonumber\\
  m_{\mu}       &=&      \pm
\frac{v_1}{\sqrt{2}}{(l_{22}-l_{23}) \over 2},   \nonumber\\
  m_{\tau}   &  =   & \pm   
 \frac{v_1}{\sqrt{2}}{(l_{22}+l_{23}) \over 2}.\ 
\eea\\
Since $m_{\mu} <<  m_{\tau}$ there has to be a  fine tuned cancellation between
$l_{22}$ and $l_{23}$ to produce  the muon mass. Hence, it is
more  natural to  consider  the symmetry  limit $l_{22}=l_{23}$  which
leads to  $m_{\mu}=0$. The Yukawa matrix which leads to the zero muon mass within the decoupled 2-3 symmetry is
\bea
 Y^L_{23} &= &\pmatrix{l_{11} & 0 &  0 \cr 0 & \frac{1}{2} l_{T} & \frac{1}{2} l_{T} \cr 0 &\frac{1}{2} l_{T} & \frac{1}{2} l_{T}}.\
\label{Yukawa-matrix-10}
\eea 

In the neutrino sector we assume that, in the symmetric limit, ${\cal {M}}_\nu$ has the general structure
\bea
{\cal {M}}_\nu & = &
\pmatrix{
a  & d & 0 \cr
d  & b & 0 \cr
0  & 0 & c
},\
\label{neutrino_symmetric}
\eea
where all the parameters are real. This can be diagonalized by the matrix
\bea
 W^\nu_{12}&=&\pmatrix{c_{12}&s_{12}&0\cr s_{12}&-c_{12}&0\cr 0&0&1\cr}
, \nonumber\\
s_{12}&\equiv & \sin{\theta_{12}}, \nonumber\\
c_{12}& \equiv & \cos{\theta_{12}},\
\label{solar}
\eea
where
\bea
\tan{2 \theta_{12}} & = & \frac{2d}{(a-b)}.\
\label{theta12}
\eea
We can then calculate $U_{PMNS}^{s}$ as
\bea
U_{PMNS}^{s}&= U^{\dagger}_\ell U_\nu, \
\label{nulepton}
\eea
with
\bea
U_\ell & = & W^l_{23}, \nonumber\\
U_\nu & = & W^\nu_{12},\
\label{tbmfactor} 
\eea
where $W^l_{23}$ and $W^\nu_{23}$ are given in Eq.~\ref{wl} and 
in Eq.~\ref{solar}.
 
This gives
\bea
U_{PMNS}^{s} & = & \pmatrix{c_{12} &s_{12}&0\cr
-\rr2 s_{12}
&\rr2 c_{12}  & \rr2\cr
\rr2 s_{12}
&- \rr2 c_{12} &
 \rr2},
\label{mns23}
\eea
which is just the $\mu-\tau$ symmetric leptonic mixing. If we require $\theta_{12}$
in Eq.~\ref{theta12}  to be independent of the parameters $a,\; b$ and $d$, then, we either have $a=b$ which leads to $\theta_{12}=\pi / 4$ and generates the BM mixing or $d=k(a-b)$ and in particular we obtain the tri-bimaximal mixing with $k=\sqrt{2}$. Hence, by choosing $a=b$, the neutrino mass matrix is given as
\bea
{\cal {M}}_\nu & = &
\pmatrix{
a  & d & 0 \cr
d  & a & 0 \cr
0  & 0 & c
}.\
\label{neusym}
\eea



\section{The Lagrangian in the Symmetric Limit}


In this section we present a simple Lagrangian that generates the mixing matrices considered in the previous section. We find that the model naturally 
generates the BM mixing though the TBM mixing can also be obtained but with introducing different flavor symmetries. Our phenomenology will be done in the scenario in which the leptonic mixing is BM in the symmetric limit.

We will use the seesaw mechanism to obtain the neutrino masses. Our model extends the SM by an additional Higgs doublet and two singlet scalars. The particle content of the model is given as
\begin{itemize}
 \item three left-handed lepton doublets $D_{\alpha_L}$, where $\alpha$ 
denotes $e,\; \mu,$ and $\tau$,
 \item three right-handed charged-lepton singlet $\alpha_R$, and
 \item three right-handed neutrino singlets $\nu_{\alpha R}$.
\end{itemize}
In the scalar sector, we employ
\begin{itemize}
 \item two Higgs doublets $\phi_j$ with vacuum expectation values, v.e.v, $\left\langle 0|\phi_j^0|0\right\rangle =\frac{v_j}{\sqrt{2}}$  and
 \item two real singlet scalar fields $\epsilon_1$  and $\epsilon_2,$ with v.e.v's $\left\langle 0|\epsilon_k^0|0\right\rangle =w_k $. 
\end{itemize}
The symmetries of the Lagrangian  are  introduced as 
\begin{eqnarray}
 Z_2^{23}&:&D_{\mu_L}\leftrightarrow -D_{\tau_L},\; \mu_R\leftrightarrow -\tau_R,\;\nu_{\mu R}\leftrightarrow -\nu_{\tau R},\nonumber \\
&&  D_{e_L} \rightarrow D_{e_L},\; e_R\rightarrow e_R ,\; \nu_{e R}\rightarrow \nu_{e R} ,\nonumber\\
&& \epsilon_{1} \rightarrow -\epsilon_{1},\; \epsilon_{2} \rightarrow \epsilon_{2},\; \phi_1 \rightarrow \phi_1 ,\; \phi_2 \rightarrow \phi_2 , \nonumber\\
  Z_2^{e}&:& \nu_{e R},\; e_R ,\; D_{e_L} ,\; \epsilon_{1},\; \epsilon_{2}, \quad     \mbox{(Change sign, and the rest of the fields remain same)}\nonumber \\
  Z_2^{D}&:& \nu_{e R},\; \nu_{\mu R},\; \nu_{\tau R},\; \phi_2 , \quad   \mbox{(Change sign, and the rest of the fields remain same)}. \nonumber \\
\label{symmetry}
\end{eqnarray}
 The most general  Lagrangian consistent with the symmetries is 
\begin{eqnarray}
{\cal L}_Y & = & \left[ y_{1} \bar{D}_{e_L}e_R  +y_2\left(\bar{D}_{\mu_L}\mu_R+\bar{D}_{\tau_L}\tau_R \right) +y_3\left(\bar{D}_{\mu_L}\tau_R+\bar{D}_{\tau_L}\mu_R \right)\right] \phi_1  \nonumber \\
 & +&\left[y_4  \bar{D}_{e_L} \nu_{eR}+ y_5 \left( \bar{D}_{\mu_L} \nu_{\mu R}+\bar{D}_{\tau_L} \nu_{\tau R}\right)  \right] \tilde{\phi_2}+h.c.,
\label{Yukawa}
\end{eqnarray}
\begin{eqnarray}
 {\cal L}_M & = & \frac{1}{2}  \left[M\nu^{T}_{e R} C^{-1} \nu_{e R}  +
M_P\nu^{T}_{\mu R} C^{-1} \nu_{\mu R}+M_P\nu^{T}_{\tau R} C^{-1} \nu_{\tau R}  \right] \nonumber\\
   &-&\frac{1}{2}y \nu^{T}_{eR} C^{-1} \left(\nu_{\mu R}\frac{(a\epsilon_{1}+b\epsilon_{2})}{\sqrt{2}}+\nu_{\tau R}\frac{(a\epsilon_{1}-b\epsilon_{2})}{\sqrt{2}}  \right)+h.c.
\label{Majorana}
\end{eqnarray}
Here, $\tilde{\phi}_i \equiv i\sigma_{2}\phi^*_i$ is the conjugate Higgs doublet
and we have chosen to work in a basis where the Dirac mass matrix for the neutrinos is diagonal.
We can simplify the Lagrangian in several ways. First, we can redefine $ a \epsilon_1 \rightarrow \epsilon_1$ and $ b \epsilon_2 \rightarrow \epsilon_2$. 
Second, to reduce the number of parameters we can impose an approximate symmetry of the Lagrangian. A $SU(3)$ symmetry where the right handed singlet fields and the left handed doublet fields transform as the $SU(3)$ triplets  leads to $y_4=y_5=y_D$. The $SU(3)$ symmetry is only satisfied by the Dirac mass term for the neutrinos and is broken by the other terms in the Lagrangian.
Third, we will require the Lagrangian to be invariant under the transformation of the right-handed charged leptons $(\mu_R \leftrightarrow - \tau_R , e_R \rightarrow - e_R , \phi_1 \rightarrow -\phi_1)$, with all other fields remaining unchanged. This symmetry requires $y_2=y_3$ leading to vanishing $\mu$ mass. The $\mu$ mass is introduced later as a symmetry breaking term. 
Finally, we will set the Majorana mass terms $M=M_P$.  
We can then rewrite the Lagrangian as
\begin{eqnarray}
{\cal L}_Y & = & \left[ y_{1} \bar{D}_{e_L}e_R +y_2\left(\bar{D}_{\mu_L}\mu_R+\bar{D}_{\tau_L}\tau_R \right) +y_2\left(\bar{D}_{\mu_L}\tau_R+\bar{D}_{\tau_L}\mu_R \right)\right] \phi_1  \nonumber \\
 & +& y_D \left[  \bar{D}_{e_L} \nu_{eR}+   \bar{D}_{\mu_L} \nu_{\mu R}+\bar{D}_{\tau_L} \nu_{\tau R}  \right] \tilde{\phi_2}+h.c.,
\label{Yukawa_new}
\end{eqnarray}
\begin{eqnarray}
 {\cal L}_M & = & \frac{1}{2} M \left[\nu^{T}_{e R} C^{-1} \nu_{e R}  +\nu^{T}_{\mu R} C^{-1} \nu_{\mu R}+\nu^{T}_{\tau R} C^{-1} \nu_{\tau R}  \right] \nonumber\\
  &-& \frac{1}{2}y \nu^{T}_{eR} C^{-1} \left(\nu_{\mu R}\frac{(\epsilon_{1}+\epsilon_{2})}{\sqrt{2}}+\nu_{\tau R}\frac{(\epsilon_{1}-\epsilon_{2})}{\sqrt{2}}  \right) +h.c. 
\label{Majorana_new}
\end{eqnarray}

The most general scalar potential $V$ that is invariant under 
$Z_2^{23}\times Z_2^{e} \times Z_2^{D}$ is given by
\bea
V & = & -\mu_1^2 \epsilon_1^2 -\mu_2^2 \epsilon_2^2  + \lambda_1 \epsilon_1^4 + \lambda_2 \epsilon_2^4  + \lambda'_1 \epsilon_1^2 \epsilon_2^2      \nonumber\\
&+& \sigma_1 \epsilon_1^2 |\phi_1|^2  + \sigma_2 \epsilon_1^2 |\phi_2|^2  + \sigma_3 \epsilon_2^2 |\phi_1|^2  + \sigma_{4} \epsilon_2^2 |\phi_2|^2 +V_{2HD}(\phi_1,\; \phi_2), 
\label{potential2}
\eea
where $V_{2HD}(\phi_1,\; \phi_2)$ is the potential of the two Higgs doublets,
\bea
V_{2HD}(\phi_1,\; \phi_2) &=& -\mu_{\phi_1}^2 \phi_1^\dagger \phi_1 -\mu_{\phi_2}^2 \phi_2^\dagger \phi_2 + \lambda_{\phi_1} (\phi_1^\dagger \phi_1)^2 + \lambda_{\phi_2} (\phi_2^\dagger \phi_2)^2 + \lambda_{\phi_{12}} \left(\phi_1^\dagger \phi_1 + \phi_2^\dagger \phi_2 \right)^2 \nonumber\\
 &+& \lambda'_{\phi_{12}} \left(\phi_1^\dagger \phi_1 - \phi_2^\dagger \phi_2 \right)^2  + \lambda_{\phi_{21}} \left( (\phi_1^\dagger \phi_1 ) (\phi_2^\dagger \phi_2 ) - (\phi_1^\dagger \phi_2 ) (\phi_2^\dagger \phi_1 ) \right) \nonumber\\
&+& \lambda'_{\phi_{21}} \left( (\phi_1^\dagger \phi_1 ) (\phi_2^\dagger \phi_2 ) + (\phi_1^\dagger \phi_2 ) (\phi_2^\dagger \phi_1 ) \right).
\eea
If we impose an additional symmetry to the above potential such as $\epsilon_1 \leftrightarrow \epsilon_2$, then the potential takes the form 
\bea
V&=& -\mu^2 \left( \epsilon_1^2 + \epsilon_2^2 \right) +  \left( \epsilon_1^2 + \epsilon_2^2 \right) \sum_{i=1}^2 \sigma_i  \phi_i^\dagger \phi_i + \lambda \left( \epsilon_1^2 + \epsilon_2^2 \right)^2 \nonumber\\
&+& \lambda' \left( \epsilon_1^2 - \epsilon_2^2 \right)^2   + V_{2HD}(\phi_1,\; \phi_2).
\label{potential}
\eea
We can parametrize the v.e.v's of the singlet scalars as follows
\bea
\left\langle 0\left|\epsilon_1\right|0\right\rangle=w \cos \gamma \quad & \mbox{and} & \quad \left\langle 0\left|\epsilon_2\right|0\right\rangle=w \sin \gamma .
\label{parametrize}
\eea 
Thus, the only term that depends on $\gamma$ is
\beq
f(\gamma)\equiv\lambda' w^4 \cos^2 2\gamma .
\eeq
By minimizing $f(\gamma)$, one gets
\bea
\cos 2\gamma &=& 0. \nonumber\\
\label{omega}
\eea
Thus 
\bea
\left\langle 0\left|\epsilon_1 \right|0\right\rangle= \left\langle 0\left|\epsilon_2 \right|0\right\rangle = \frac{w}{\sqrt{2}}.
\eea
By minimizing the above potential one can find the parameter $w$ and the v.e.v's of the two Higgs doublets which are nonzero and different in the symmetric limit
\bea
v_1 &=& \sqrt{\frac{\alpha_1}{2\beta_1}},  \nonumber\\
v_2 &=& \sqrt{\frac{\alpha_2}{2\beta_2}},
\eea
where
\bea
\alpha_1 &=& 4 \lambda (\lambda_{\phi_{12}} \mu_{\phi_{1}}^2 + \lambda_{\phi_2} \mu_{\phi_1}^2 - \lambda'_{\phi_{21}} \mu_{\phi_2}^2 - \lambda_{\phi_{12}} \mu_{\phi_2}^2 + \lambda'_{\phi_{12}} (\mu_{\phi_1}^2 + \mu_{\phi_2}^2)) -2 \lambda_{\phi_{12}} \mu^2 \sigma_1 \nonumber\\
&-&  2 \lambda_{\phi_2} \mu^2 \sigma_1 +  2 \lambda'_{\phi_{21}} \mu^2 \sigma_2 + 
 2 \lambda_{\phi_{12}} \mu^2 \sigma_2 + \mu_{\phi_2}^2 \sigma_1 \sigma_2 - \mu_{\phi_1}^2 \sigma_2^2 - 
 2 \lambda'_{\phi_{12}} \mu^2 (\sigma_1 + \sigma_2), \nonumber\\
\beta_1  &=& 4 \lambda (-\lambda'^2_{\phi_{21}} - 
    2 \lambda'_{\phi_{21}} \lambda_{\phi_{12}} + \lambda_{\phi_1} 
\lambda_{\phi_{12}} + \lambda_{\phi_1} \lambda_{\phi_2} + \lambda_{\phi_{12}} \lambda_{\phi_2} + \lambda'_{\phi_{12}} (2 \lambda'_{\phi_{21}} + \lambda_{\phi_1}+4 \lambda_{\phi_{12}} \nonumber\\ 
&+&   \lambda_{\phi_2})) - \lambda_{\Phi_{12}} \sigma_1^2 - \lambda_{\phi_2} \sigma_1^2 + 
 2 \lambda'_{\phi_{21}} \sigma_1 \sigma_2 + 
 2 \lambda_{\phi_{12}} \sigma_1 \sigma_2 - \lambda_{\phi_1} 
\sigma_2^2 - \lambda_{\phi_{12}} \sigma_2^2 - \lambda'_{\phi_{12}} (\sigma_1 + \sigma_2)^2 , \nonumber\\
\alpha_2 &=& 4 \lambda (\lambda_{\phi_{12}} \mu_{\phi_{2}}^2 + \lambda_{\phi_1} \mu_{\phi_2}^2 - \lambda'_{\phi_{21}} \mu_{\phi_1}^2 - \lambda_{\phi_{12}} \mu_{\phi_1}^2 + \lambda'_{\phi_{12}} (\mu_{\phi_2}^2 + \mu_{\phi_1}^2)) -2 \lambda_{\phi_{12}} \mu^2 \sigma_2 \nonumber\\
&-&  2 \lambda_{\phi_1} \mu^2 \sigma_2 +  2 \lambda'_{\phi_{21}} \mu^2 \sigma_1 + 
 2 \lambda_{\phi_{12}} \mu^2 \sigma_1 + \mu_{\phi_1}^2 \sigma_2 \sigma_1 - \mu_{\phi_2}^2 \sigma_1^2 - 
 2 \lambda'_{\phi_{12}} \mu^2 (\sigma_2 + \sigma_1), \nonumber\\
\beta_2  &=& 4 \lambda (-\lambda'^2_{\phi_{21}} - 
    2 \lambda'_{\phi_{21}} \lambda_{\phi_{12}} + \lambda_{\phi_2} 
\lambda_{\phi_{12}} + \lambda_{\phi_2} \lambda_{\phi_1} + \lambda_{\phi_{12}} \lambda_{\phi_1} + \lambda'_{\phi_{12}} (2 \lambda'_{\phi_{21}} + \lambda_{\phi_2}+4 \lambda_{\phi_{12}} \nonumber\\ 
&+&   \lambda_{\phi_1})) - \lambda_{\Phi_{12}} \sigma_2^2 - \lambda_{\phi_1} \sigma_2^2 + 
 2 \lambda'_{\phi_{21}} \sigma_2 \sigma_1 + 
 2 \lambda_{\phi_{12}} \sigma_2 \sigma_1 - \lambda_{\phi_2} 
\sigma_1^2 - \lambda_{\phi_{12}} \sigma_1^2 - \lambda'_{\phi_{12}} (\sigma_2 + \sigma_1)^2 . \nonumber\\
\eea
Also, the parameter $w$ can simply be written as follows
\beq
w^2 = \frac{\mu^2-(\sigma_1 |v_1|^2 + \sigma_2  |v_2|^2)}{2 \lambda},
\eeq
which shows that the v.e.v of the singlet scalars is independent of $(v_1,\; v_2)$ when $\sigma_1 = \sigma_2 = 0$.

The explicit form of the Yukawa matrix, $Y^L_{23}$, and the Dirac neutrino mass matrix can be written from the Lagrangian (\ref{Yukawa}) as follows
\begin{eqnarray}
 Y^L_{23} & = &\frac{v_1}{\sqrt{2}} \pmatrix{ y_1   &  0  & 0   \cr  
                                              0   &  y_2  &  y_2   \cr  
                                              0   &  y_2  &  y_2   }, \
\label{PMNS-Yukawa}
\end{eqnarray}
\begin{equation}
 M_D=\mbox{diag}(A,A,A), \quad \mbox{with}\; A= y \frac{v_2}{\sqrt{2}}.
\end{equation}
 Also, the Majorana mass matrix can be obtained from Eq.~(\ref{Majorana}) as follows
\beq
M_R  =  \pmatrix{M & - v_w &  0  \cr -v_w   &  M   &  0 \cr  0  & 0  & M },  
\label{numatrix1}
\eeq
with $ v_w=yw$.
Using the seesaw formula \cite{seesaw-1}, the neutrino mass matrix is given as
\bea
{\cal M}_\nu & = & -M^T_D M^{-1}_R M_D.
\label{seesaw}
\eea
Then ${\cal M}_\nu$ has the structure 
\bea
{\cal M}_\nu & = & \pmatrix{X & G & 0 \cr G & X & 0 \cr 0 & 0 & Z},
\label{mass-seesaw}
\eea
where 
\bea
X&=&-\frac{A^{2}M}{M^2 - v_w^2},\; G=-\frac{A^{2} v_w }{M^2 - v_w^2},\; Z=-\frac{A^{2}}{M}.
\label{numatrix2}
\eea
By diagonalizing Eq.~\ref{mass-seesaw}, we  obtain the neutrino masses as
\begin{eqnarray}
 m_1 & = & -\frac{A^2}{M +v_w},\nonumber \\
 m_2 & = & -\frac{A^2}{M -v_w},\nonumber \\
 m_3 & = & -\frac{A^{2}}{M}.
\end{eqnarray}

Note that from the above equations one can estimate the scale of the v.e.v, $v_2$, of the second Higgs doublet $\phi_2$. As the absolute neutrino masses are in the eV scale, therefore,  $v_2$ has to be in the MeV scale if the see-saw scale $(M)$ is in the TeV range. The mass relations satisfy the relation
\beq
\frac{1}{m_1}+\frac{1}{m_2}=\frac{2}{m_3}.
\eeq
Similar relations among the masses are discussed in Ref.~\cite{Barry:2010yk}.
 We can use the above sum-rule to obtain an upper limit for the heaviest mass, $|m_3 |\leqslant \frac{2|m_1| |m_2|}{||m_1| + |m_2||}$ for the normal hierarchy or $|m_2 |\leqslant \frac{|m_1| |m_3|}{|2|m_1| - |m_3||}$ for the inverted hierarchy.


 \section{Symmetry Breaking}


 The breaking  of the flavor symmetries in the charged lepton  and the neutrino
 sectors will cause deviation from the BM form, and we study these deviations in this section.

 \subsection{Charged Lepton Sector}
In the charged lepton sector we break the decoupled $2-3$ symmetry by adding the following higher dimensional terms
\beq
O_1=c y_2\bar{D}_{\mu_L}\mu_R \phi_1 \frac{\phi_1^\dagger \phi_1}{\Lambda^2}, 
\label{high-term-1}
\eeq
and
\beq
O_2= y' \left( \bar{D}_{e_L}\mu_R - \bar{D}_{e_L}\tau_R + \bar{D}_{\mu_L}e_R - \bar{D}_{\tau_L}e_R\right) \phi_1 \frac{\phi_1^\dagger \phi_1}{\Lambda^2}. 
\label{high-term-2}
\eeq
The operator $O_2$ breaks the decoupled $2-3$ symmetry, $Z_2^{23} \times Z_2^e$, but is still $2-3$ symmetric. The operator $O_1$ explicitly breaks the $2-3$ symmetry, $Z_2^{23}$, and generates the muon mass.
To generate explicit $2-3$ breaking we have introduced the higher dimensional operator in the position of the muon field, 2-2 element, in the Yukawa matrix which is the most straightforward way to generate the muon mass. Introducing this operator in the 3-3 position  generates the same numerical solutions for the correction angles. But introducing it in the 2-3 or 3-2 positions does not generate physical values for the mixing angles. 
Even introducing $2-3$ symmetric terms in (2-2, 3-3) or (2-3, 3-2)  generates either unphysical mixing angles or gives very large correction mixing angles that do not lead to successful phenomenology. 

 In the presence of the higher dimensional terms the charged lepton Yukawa matrix has the following form
\bea
 Y^L &= &\pmatrix{l_{11} & l_{12} &  -l_{12} \cr l_{12} & \frac{1}{2}{l_{T}}(1+2\kappa_l) & \frac{1}{2}{l_{T} }\cr -l_{12}
&\frac{1}{2}{l_{T}} & \frac{1}{2}{l_{T}}},\
\label{Yukawa-matrix-6}
\eea
with $\kappa_l=c v_1^2 /2 \Lambda^2 $ and $l_{12}=y'v_1^3 / 2\sqrt{2}\Lambda^2$ after the Higgs field gets it's v.e.v. Three relations can be obtained among the $Y^L$ matrix elements
\bea
Y^L_{12} & = & -Y^L_{13}, \nonumber\\
Y^L_{23} & = &  Y^L_{33}, \nonumber\\
Y^L_{22} & = & (1+2 \kappa_l)Y^L_{23}.
\label{relations-YL} 
\eea

We can solve for the unitary matrix, $U_l$, that diagonalizes  $Y^L$ in Eq.~\ref{Yukawa-matrix-6}.
We write,
\beq
U_l=W^l_{23} R^l_{23}R^l_{13}R^l_{12},
\label{triplet}
\eeq 
where  
 \bea
 R_{12}^{l} & = & \pmatrix{c_{12l} & s_{12l}
&  0  \cr  -s_{12l}  &  c_{12l}   &  0\cr 0  &
0& 1}, \nonumber\\
c_{12l} & = & \cos { \theta_{12l}} ; s_{12l}  =  \sin { \theta_{12l}},\
\label{12l}
\eea
\bea
 R_{13}^{l} & = & \pmatrix{c_{13l} & 0
&  s_{13l} e^{-i \delta} \cr  0  &  1   &  0\cr -s_{13l} e^{i \delta}  &
0& c_{13l}}, \nonumber\\
c_{13l} & = & \cos { \theta_{13l}} ; s_{13l}  =  \sin { \theta_{13l}},\
\label{13l}
\eea
\bea
R^l_{23} & = & \pmatrix{1 & 0
&  0  \cr  0  &  c_{23l}   &  s_{23l}\cr  0  &-s_{23l}  &
c_{23l}}, \nonumber\\ 
c_{23l} & = & \cos { \theta_{23l}} ; s_{23l}  =  \sin { \theta_{23l}}. \
\label{23d}
 \eea
The Yukawa matrix, $Y^L$, can be written as
\beq
Y^L=U_l Y^L_d U_l^{\dagger},
\label{diagonalization}
\eeq
with
\bea
 Y^L_{d} & = &\begin{pmatrix}{l_{e} & 0 &  0 \cr 0 & l_{\mu} & 0\cr 0
&0 & l_{\tau}}\end{pmatrix}.
\eea
Applying the relations in Eq.~\ref{relations-YL} to the $Y^L$ matrix elements in Eq.~\ref{diagonalization} using Eq.~\ref{triplet}, one can solve for the corrections of the mixing angles. Two ways can be used to find the angles, analytically or numerically.   Solving for the mixing angles analytically, see details in appendix~\ref{app:charged}, can determine the size of the Yukawa matrix parameters in Eq.~\ref{Yukawa-matrix-6}
\bea
z_\mu &\equiv &{ m_{\mu} \over m_{\tau}}, \nonumber\\
\kappa_l &= &z z_\mu , \nonumber\\
l_{12} &\approx &\sqrt{\frac{z_\mu}{2}} (l_e - l_\mu), \nonumber\\
l_{T} & \approx & (l_{\tau}-l_{\mu}) (1-\frac{1}{2}(z z_\mu)^2).
\eea
with $z\sim 2$. It is interesting to note that 
\beq
\kappa = \frac{c v_1^2}{2 \Lambda^2} = z z_\mu
\eeq
which fixes $\Lambda \sim $ TeV. We assume that the charged lepton corrections are ``CKM-like'', i.e. 
\beq
\sin \theta_{12l}\approx\lambda , \quad \sin \theta_{23l}\approx A \lambda^2 , \quad \sin \theta_{13l}\approx B\lambda^3 ,
\label{CKM-like}
\eeq
where $A$ and $B$ are real and of order one and $\lambda$ is the sine of the Cabibbo angle, $\lambda = \sin \theta_C \simeq 0.227$.  We present numerical solutions to the relations in Eq.~\ref{relations-YL} for various $z$ values that produce the pattern in Eq.~\ref{CKM-like}. In our calculations we assume $\delta = \pi$,

\begin{itemize}
	\item For $z=2.0$: $s_{12l}\approx\pm 0.34,  \;s_{13l}\approx\pm 0.0011,\; s_{23l}\approx - 0.059$,
	\item For $z=2.06$: $s_{12l}\approx\pm 0.3,  \;s_{13l}\approx\pm 0.001,\; s_{23l}\approx - 0.061$,
	\item For $z=2.2$: $s_{12l}\approx\pm 0.2,  \;s_{13l}\approx\pm 0.00075,\; s_{23l}\approx - 0.065$,
\end{itemize}

We expand the angles in Eq.~\ref{PMNS2} as
\begin{equation}
s_{13} = \frac{r}{\sqrt{2}}, \ \ s_{12} = \frac{1}{\sqrt{2}}(1+s), \ \ s_{23} = \frac{1}{\sqrt{2}}(1+a),
\label{rsa}
\end{equation}
where the three real parameters
$r,s,a$  describe the deviations of the reactor, solar, and
atmospheric angles from their bimaximal values. We  use
global fits of the conventional mixing parameters $(s,\; a)$ \cite{Schwetz:2008er} that can be translated into $3\sigma$ ranges and the mixing parameter $r$ with $2.5 \sigma$ significance (90\% C.L.) \cite{Abe:2011sj}
\begin{eqnarray}
0.12<r<0.39, \ -0.29<s<-0.14, \ -0.15<a<0.16. 
\label{ranges}
\end{eqnarray}
To first order in  $r,s,a$, the lepton mixing matrix can be written as,
\begin{eqnarray}
U \approx
\left( \begin{array}{ccc}
\frac{1}{\sqrt{2}}(1-s)  & \frac{1}{\sqrt{2}}(1+s) & \frac{1}{\sqrt{2}}re^{-i\delta } \\
-\frac{1}{2}(1+s-a + \frac{r}{\sqrt{2}}e^{i\delta })  & \frac{1}{2}(1-s-a- \frac{r}{\sqrt{2}}e^{i\delta })
& \frac{1}{\sqrt{2}}(1+a) \\
\frac{1}{2}(1+s+a- \frac{r}{\sqrt{2}}e^{i\delta })  & -\frac{1}{2}(1-s+a+ \frac{r}{\sqrt{2}}e^{i\delta })
 & \frac{1}{\sqrt{2}}(1-a)
\end{array}
\right),
\label{MNS1}
\end{eqnarray}
which is similar to the parametrization in Ref.~\cite{king} with the TBM mixing. We have assumed that $\delta= \pi$ where the present data prefers a negative value for $s$ \cite{king} and $r$ is positive, in our discussion we do not consider CP violation. Now, we can write the parameters $(r,\; s,\; a)$ in terms of the elements of the mixing matrix 
\bea
s & = & -1+\sqrt{2} U_{12}, \nonumber\\
r & = & \sqrt{2}(1+s-a+2 U_{21}), \nonumber\\
a & = & -1+\sqrt{2} U_{23}.
\label{Full-Equations2}
\eea 
From the details in appendix A, one obtains
\bea
 s & \approx & -\frac{1}{\sqrt{2}}(s_{12l}+s_{13l}), \nonumber\\
 r & \approx & s_{12l}-s_{13l}, \nonumber\\
 a & \approx & -s_{23l}. 
\label{relations}
\eea

From the above equations one can get the deviation parameters as follows 
\begin{itemize}
	\item For $z=2.0$:  $s \approx -0.24,\; r \approx 0.34,\; a \approx  0.059$,
	\item For $z=2.06$: $s \approx -0.21,\; r \approx 0.30,\; a \approx  0.061$,
	\item For $z=2.2$:  $s \approx -0.14,\; r \approx 0.20,\; a \approx  0.065$.
\label{numbericalresults2}
\end{itemize}
The above results demonstrate that the contributions from the charged lepton 
sector can accommodate the T2K data of $\theta_{13}$ as well as the other mixing angles.


\subsection{Neutrino Sector}


In this section we consider deviations of the BM mixing from the neutrino sector. We maintain the invariance of the Majorana Lagrangian  under the symmetry group in Eq.~\ref{symmetry} and  generate the deviation from the BM matrix by breaking the $\epsilon_1 \leftrightarrow \epsilon_2$ symmetry in Eq.~\ref{potential}  by introducing the most general dimension four symmetry breaking terms in the potential
\beq
 \left( \epsilon_1^2 - \epsilon_2^2 \right) \sum_{i=1}^2 \sigma'_i \phi_i^\dagger \phi_i + \varrho \left( \epsilon_1^2 - \epsilon_2^2 \right) \left( \epsilon_1^2 + \epsilon_2^2 \right).
\footnote{ The most general symmetry breaking terms can be expressed  
in terms of the form in Eq.~\ref{sym-breaking} and symmetry conserving terms that can be absorbed in the symmetric potential.}
\label{sym-breaking}
\eeq 
We  require that all terms in the symmetry breaking potential are of the same size which results in $ \rho \sim { v^2 \over w^2}\sigma'_i $ where $v$ is the electroweak v.e.v with
$ v^2=v_1^2+v_2^2$ and $\omega$ is the scale of the v.e.v's of the singlet scalars.
Thus, the potential is
\bea
V&=& -\mu^2 \left( \epsilon_1^2 + \epsilon_2^2 \right) +  \left( \epsilon_1^2 + \epsilon_2^2 \right) \sum_{i=1}^2  \sigma_i \phi_i^\dagger \phi_i  + \lambda \left( \epsilon_1^2 + \epsilon_2^2 \right)^2 + \lambda' \left( \epsilon_1^2 - \epsilon_2^2 \right)^2 \nonumber\\
&+&  \left( \epsilon_1^2 - \epsilon_2^2 \right) \sum_{i=1}^2  \sigma'_i \phi_i^\dagger \phi_i + \varrho \left( \epsilon_1^2 - \epsilon_2^2 \right) \left( \epsilon_1^2 + \epsilon_2^2 \right) + V_{2HD}(\phi_1,\; \phi_2).
\label{potential-10}
\eea
Now, parameterizing the v.e.v's as in Eq.~\ref{parametrize} and minimizing the potential leads to
\bea
\cos 2\gamma &=& -\frac{\varrho w^2 +(\sigma'_1  |v_1|^2 + \sigma'_2  |v_2|^2)}{2\lambda' w^2  }, \nonumber\\
w^2 &=& \frac{2\lambda' (\mu^2 - (\sigma_1 |v_1|^2 + \sigma_2  |v_2|^2)) + \varrho (\sigma'_1  |v_1|^2 + \sigma'_2  |v_2|^2)  }{4 \lambda \lambda' - \varrho^2}.
\label{omega-2}
\eea
Keeping in mind the size of the various co-efficients in the symmetry breaking potential discussed above, we find that $\cos 2\gamma\approx 0$ up to corrections of order ${v^2 \over \omega^2}$. We assume that $w$ is in the TeV scale and with $v$ in the EW scale the symmetry breaking corrections are of the right size to explain the experimental numbers.

We  shift the v.e.v's of the two singlet scalars ($w_1\neq w_2$) up to the first order of the symmetry breaking parameter. Then, the Majorana neutrino mass matrix in Eq.~\ref{numatrix1}  takes the form  
\beq
M_R  =  \pmatrix{M & - v_{w p} &  - v_{w n}  \cr -v_{w p}   &  M   &  0 \cr  - v_{w n}  & 0  & M  } ,  \
\label{numatrixbreak}
\eeq
where
\bea
v_{w p}&=&\frac{y}{\sqrt{2}}(w_1+w_2),\nonumber\\
v_{w n}&=&\frac{y}{\sqrt{2}}(w_1-w_2). 
\label{pn}
\eea
We  write the v.e.v's of the singlet scalars after symmetry breaking
 as  
\bea
w_1 &= & \frac{w+\rho_1 }{\sqrt{2}},\nonumber \\
w_2 & = & \frac{w+\rho_2}{\sqrt{2}},
\eea
where $\rho_1$ and $\rho_2$ are small quantities and
\beq
\rho_1 =-\rho_2 = \frac{w \tau}{2}.
\eeq
Up to the first order of  the symmetry breaking parameter $\tau$,
\beq
\tau \equiv -\frac{\varrho w^2 +(\sigma'_1  |v_1|^2 + \sigma'_2  |v_2|^2)}{2\lambda' w^2  },
\eeq
one gets
\bea
v_{w p}&=&v_{w},\nonumber\\
v_{w n}&=&\frac{\tau}{2} v_w . 
\label{pn2}
\eea
It turns out that breaking the $\epsilon_1 \leftrightarrow \epsilon_2 $ symmetry to generate different v.e.v's for the singlet scalars is  not sufficient to break the almost degeneracy of $(m_1 ,\; m_2)$ to satisfy the squared mass difference measurements. Therefore, we  introduce an additional term in the 
Lagrangian which is consistent with the symmetries of the Lagrangian,
\beq
M_1 \left[ \nu^{T}_{\mu R} C^{-1} \nu_{\mu R}+\nu^{T}_{\tau R} C^{-1} \nu_{\tau R}  \right].
\eeq
Thus 
\beq
M_R  =  \pmatrix{M & - v_{w} &  - \frac{\tau}{2} v_w   \cr -v_{w}   &  M'   &  0 \cr  - \frac{\tau}{2} v_w   & 0  & M'  } ,  \
\label{numatrixbreak-2}
\eeq
where $M'= M + M_1$.

 The neutrino mass matrix in Eq.~\ref{mass-seesaw}  changes to be 
\bea
{\cal M}_\nu & = & \pmatrix{X' & G' & P' \cr G' & Y' & W' \cr P' & W' & Z'},
\label{mass-seesaw-break}
\eea
where 
\bea
X' & = &  -\frac{4 A^2 M'}{4M M' -v_w^2 (4+\tau^2)} ,   \nonumber \\
Y' & = &  -\frac{A^2 (4 M M' - v_w^2 \tau^2)}{M'(4M M' -v_w^2 (4+\tau^2))} ,   \nonumber \\
Z' & = &  -\frac{4A^2 (M M' - v_w^2 )}{M'(4M M' -v_w^2 (4+\tau^2))} ,   \nonumber \\
G' & = &  -\frac{4A^2 v_w}{4M M' -v_w^2 (4+\tau^2)} ,   \nonumber \\
P' & = &  -\frac{2A^2 v_w \tau}{4M M' -v_w^2 (4+\tau^2)}  ,    \nonumber \\
W' & = &  -\frac{2 A^2 v_w^2 \tau}{M'(4M M' -v_w^2 (4+\tau^2))} .   \nonumber \\
\label{mass-seesaw-break2}
\eea

By diagonalizing Eq.~\ref{mass-seesaw-break}, one gets the mass eigenvalues
\bea
m_1 &=& - \frac{2A^2 \left( (M+M')-\sqrt{M^2 -2MM'+M'^2+v_w^2 (4+\tau^2)}  \right)}{4MM'-v_w^2 (4+ \tau^2)},\nonumber\\
m_2 &=& - \frac{2A^2 \left( (M+M')+\sqrt{M^2 -2MM'+M'^2+v_w^2 (4+\tau^2)}  \right) }{4MM'-v_w^2 (4+ \tau^2)},\nonumber\\
m_3 &=& - \frac{A^2}{M'}.
\label{masses}
\eea

Now, we can  diagonalize the mass matrix in Eq.~\ref{mass-seesaw-break} using the unitary matrix
 $U_{\nu}=W^{\nu}_{12}R_{23}^{\nu}R_{12}^{\nu}$ with,
\bea
 R_{12}^{\nu} & = & \pmatrix{ c_{12\nu} & s_{12\nu} & 0 \cr 
                              -s_{12\nu} & c_{12\nu} & 0 \cr
                              0 & 0 & 1 }, \nonumber\\
c_{12\nu} & = & \cos { \theta_{12\nu}} ; s_{12\nu}  =  \sin { \theta_{12\nu}},\nonumber\\
R_{23}^{\nu} & = & \pmatrix{ 1 & 0 & 0 \cr 
                              0 & c_{23\nu} & s_{23\nu} \cr
                              0 & -s_{23\nu} & c_{23\nu} }, \nonumber\\
c_{23\nu} & = & \cos { \theta_{23\nu}} ; s_{23\nu}  =  \sin { \theta_{23\nu}}.
\eea

The mass matrix elements in Eq.~\ref{mass-seesaw-break2} satisfy the two relations
\begin{eqnarray}
 X'(Z'-Y')&=&P'^2-G'^2, \nonumber\\
 G' P' (Z'-Y')&=& W'(P'^2-G'^2).
\end{eqnarray}
By applying the above relations to the matrix elements of 
\beq
{\cal {M}}_\nu = U_\nu {\cal {M}}_\nu^d U_\nu^\dagger ,
\eeq
one can obtain the mixing angles
\bea
s_{23\nu} &=&  \sqrt{\frac{2  m_1 (m_2 - m_3)}{m_2 (m_1 - m_3)}}, \nonumber\\
s_{12\nu} &=&\sqrt{\frac{- m_1 m_2 + 2 m_1 m_3 - m_2 m_3 }{2 m_3 (m_1 - m_2) }}. 
\label{nu-correction-angles}
\eea

Eventually, we obtain the elements of the lepton mixing matrix $U_{PMNS}=U_{l}^{\dag}U_{\nu}$ with $U_\ell  =  W^l_{23} R^l_{23} R^l_{13} R^l_{12}$ and $U_{\nu}=W^{\nu}_{12}R_{23}^{\nu}R_{12}^{\nu}$. The deviation parameters ($s,\; r,\; a$) can be obtained from Eq.~\ref{Full-Equations2} as follows
\begin{eqnarray}
 s & \approx & -\frac{1}{\sqrt{2}}(s_{12l}+s_{13l})+s_{12\nu}, \nonumber\\
 r & \approx & s_{12l}-s_{13l}-s_{23\nu}, \nonumber\\
 a & \approx & -s_{23l}+\frac{1}{\sqrt{2}}s_{23\nu}. \
 \label{final}
\end{eqnarray}


\section{Numerical Results}


From the neutrino mass matrix (\ref{neusym}), one  observes that in the degenerate case, when $m_1\approx m_2\approx m_3$, $a\approx c$, $d \approx 0$ which means that the neutrino mass matrix is already diagonalized as ${\cal {M}}_{\nu}\approx \mbox{diag}\left(a,a,a\right)$. That means the lepton mixing matrix does not include a contribution from the neutrino sector, and the resultant leptonic mixing is inconsistent with the experimental data. Thus, in the symmetric limit our model excludes the  case of the degenerate neutrino masses. Even, after symmetry breaking, the degenerate case in Eq.~\ref{masses} leads to vanishing 
the v.e.v's of the singlet scalar fields which does not lead to successful phenomenology.

The numerics goes as following; we choose  masses $(m_1 , m_2 , m_3)$ which satisfy the experimental values of the squared mass differences
\bea
\Delta m_{21}^2 &=& m_2^2 - m_1^2 = (7.59\pm 0.20)\times 10^{-5} eV^2 ,\nonumber\\
\Delta m_{32}^2 &=& |m_3^2 - m_2^2| = (2.43 \pm 0.13)\times 10^{-3} eV^2. 
\eea
We substitute those mass  values in ($r, s, a$) in Eq.~\ref{Full-Equations2},
 using $(s_{12\nu},\; s_{23\nu})$  given in Eq.~\ref{nu-correction-angles} and $(s_{12l},\; s_{23l},\; s_{13l})$  in sec.~(4.1). If the results satisfy the experimental constraints in Eq.~\ref{ranges}, we plot the possible values of the absolute masses and the mixing angles. By using Eq.~\ref{masses}, we calculate values for the Lagrangian parameters $(v_w,\; A,\; M,\; M')$ which generate the values of the absolute masses  obtained from the graphs.  From the graphs, one find that $(v_w,\; M,\; M')$ are obtained in the TeV scale and $A$ in the MeV range.

Three mass-dependent neutrino observables are probed in different types of experiments. The sum of absolute neutrino masses $m_{cosm}\equiv\Sigma m_i$  is probed in cosmology, the kinetic electron neutrino mass in  beta decay $(M_\beta)$ is probed in direct search for neutrino masses, and the effective mass $(M_{ee} )$ is probed in neutrinoless double beta decay experiments with the decay rate for the process $\Gamma \propto M_{ee}^2$. In terms of the ``bare'' physical parameters $m_i$ and $U_{\alpha i}$, the observables are given by \cite{Barry:2010yk}
\bea
\Sigma m_i &=& |m_1|+|m_2|+|m_3|, \nonumber\\
M_{ee}  &=& ||m_1||U_{e1}|^2 +|m_2||U_{e2}|^2 e^{i\phi_1}+|m_3||U_{e3}|^2 e^{i\phi_2}|, \nonumber\\
M_\beta &=& \sqrt{|m_1|^2 |U_{e1}|^2 +|m_2|^2 |U_{e2}|^2 +|m_3|^2 |U_{e3}|^2}.
\eea
In our analysis we ignore the Majorana phases $(\phi_1 ,\; \phi_2)$ and plot  $ M_{\beta}$ versus $\Sigma m_i $ and $M_{ee}$ versus $m_{light}$, where $m_{light}$ is the lightest neutrino mass.

\begin{figure}[b]
	\centering
		\includegraphics[width=7.3cm]{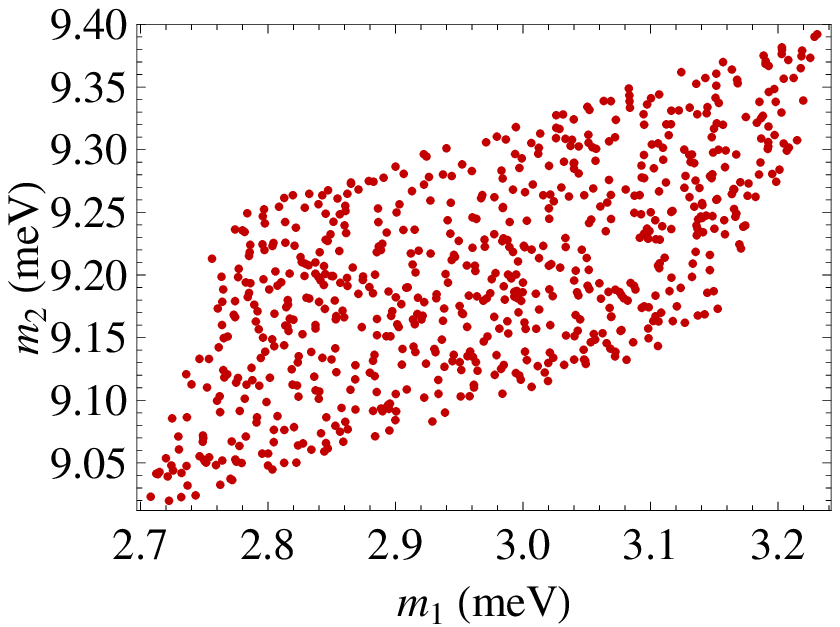}
		\includegraphics[width=7.3cm]{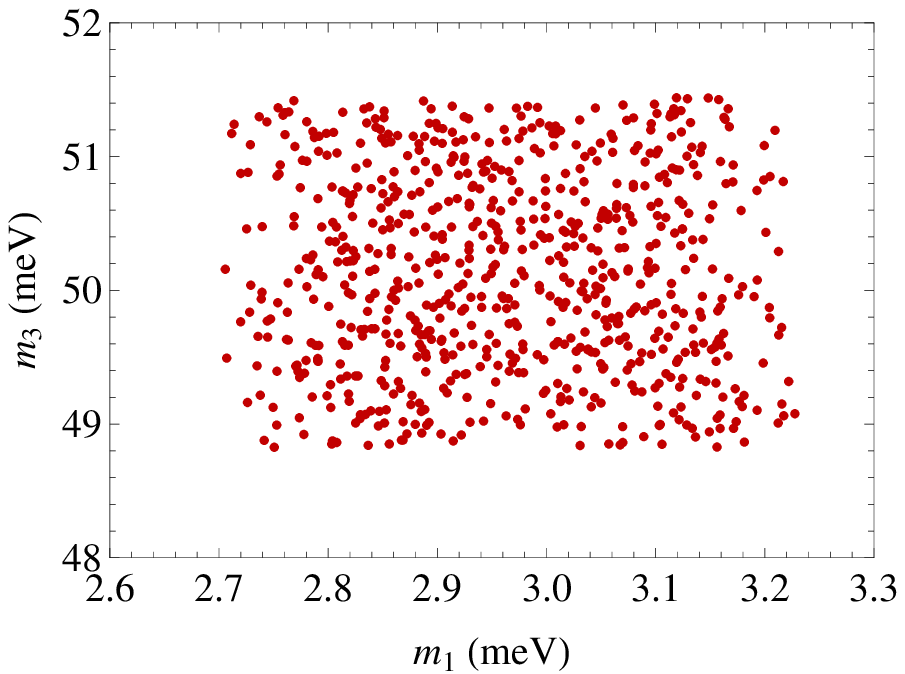}
		\includegraphics[width=7.3cm]{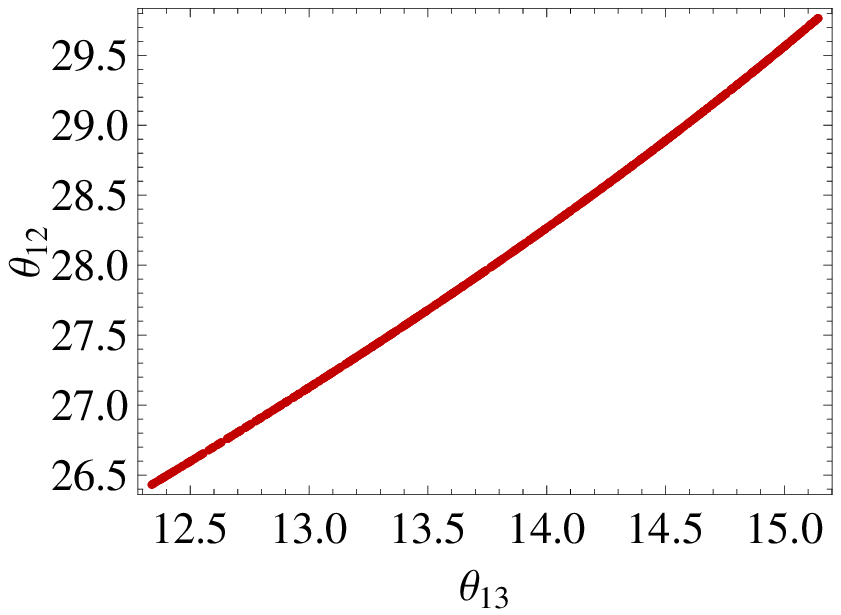}
		\includegraphics[width=7.3cm]{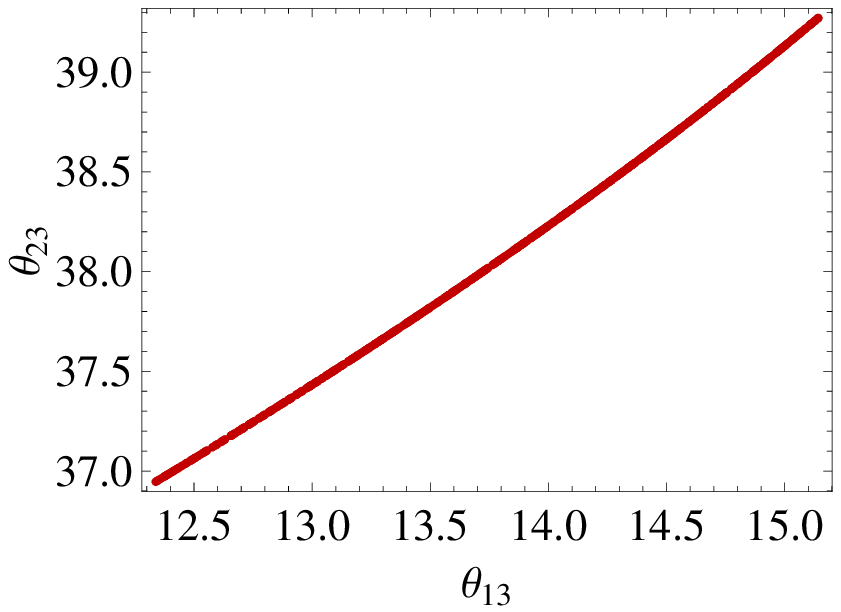}
		\includegraphics[width=7.3cm]{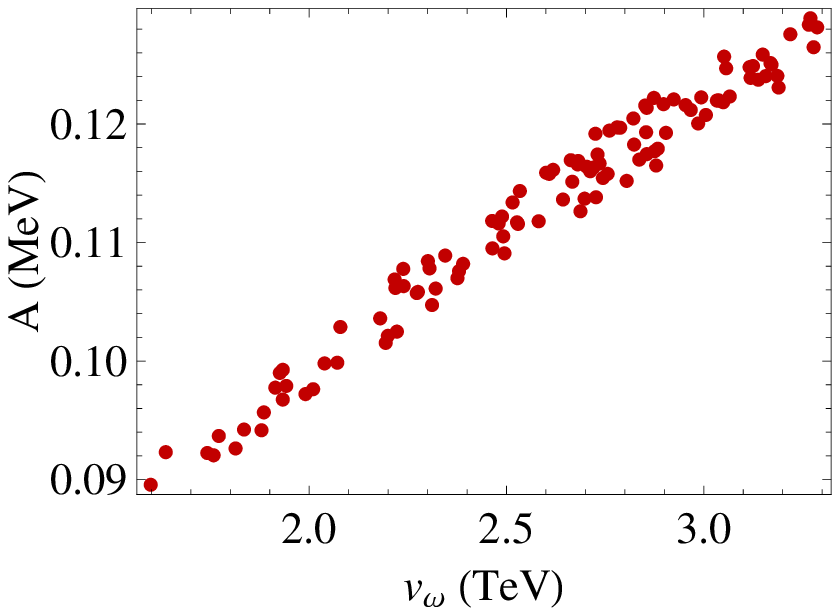}
		\includegraphics[width=7.3cm]{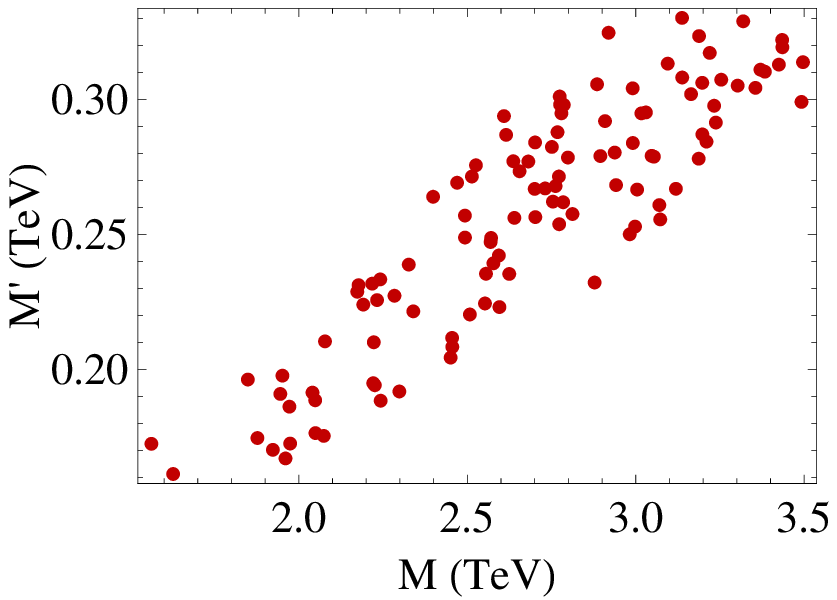}
		\includegraphics[width=7.3cm]{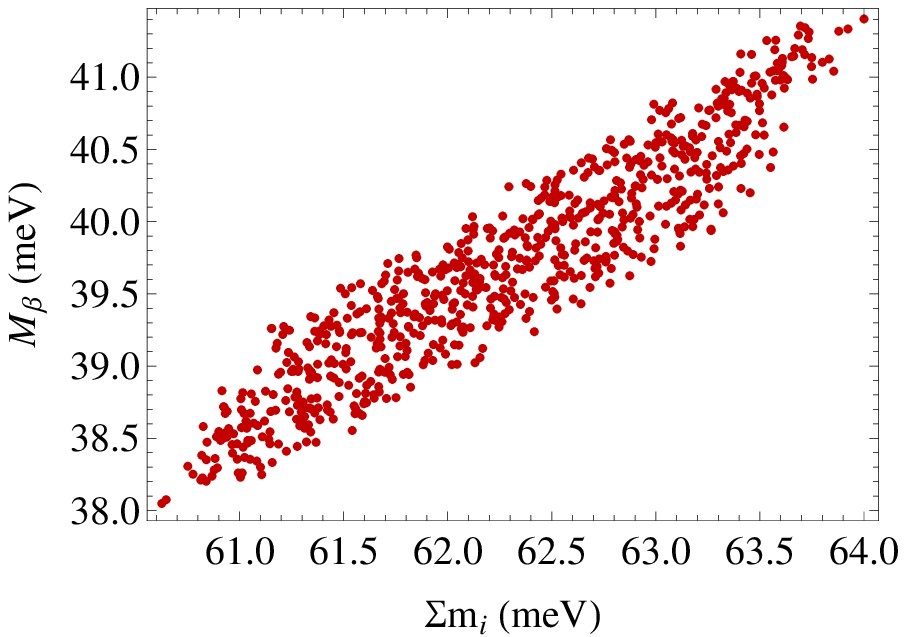}
		\includegraphics[width=7.3cm]{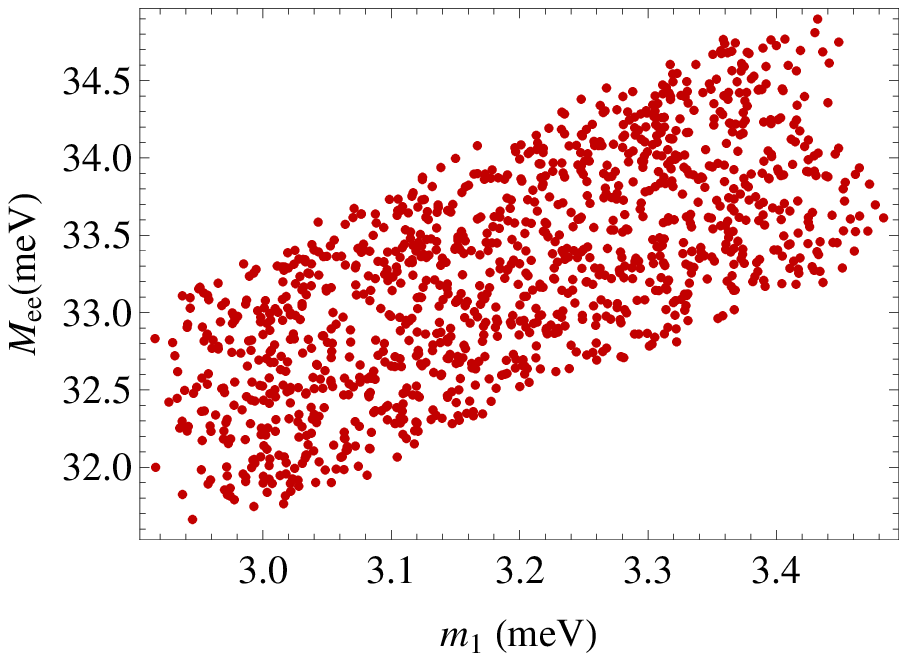}
	\caption{Scatter plots for $z=2.0$ with $s_{12l}\approx  -0.34,\; s_{13l}\approx  -0.0011$, and  $s_{23l} \approx -0.059$. In the neutrino sector, we assume that $\tau=0.1 $. (meV $\equiv 10^{-3}$ eV)}
	\label{plot1}
\end{figure}

\begin{figure}[b]
	\centering
		\includegraphics[width=7.3cm]{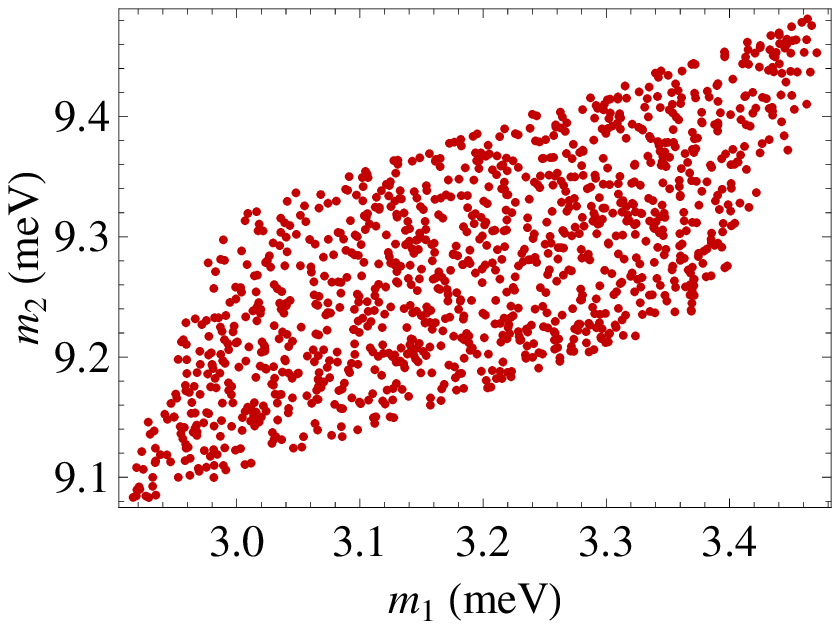}
		\includegraphics[width=7.3cm]{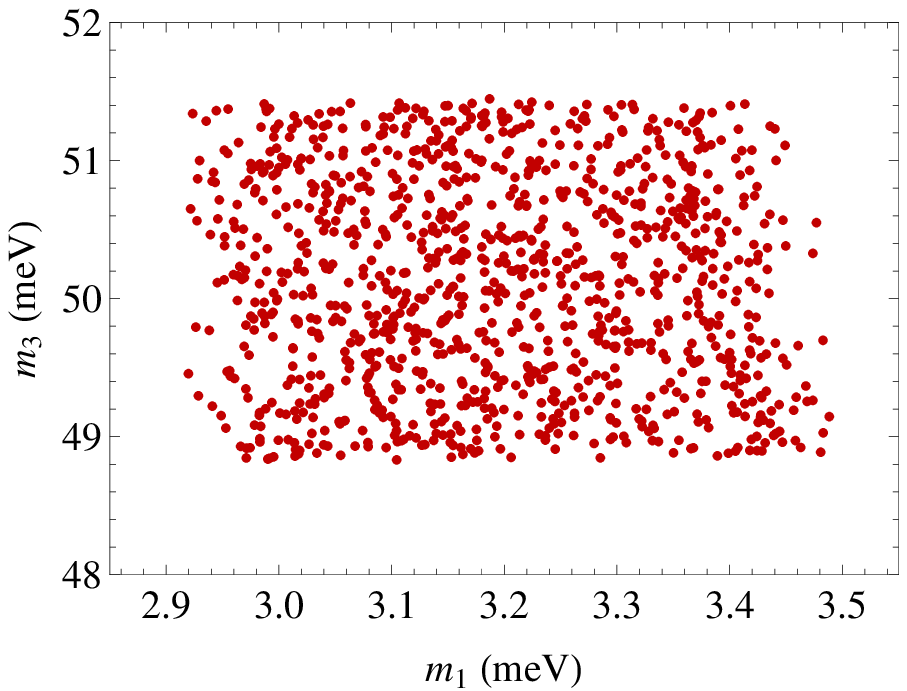}
		\includegraphics[width=7.3cm]{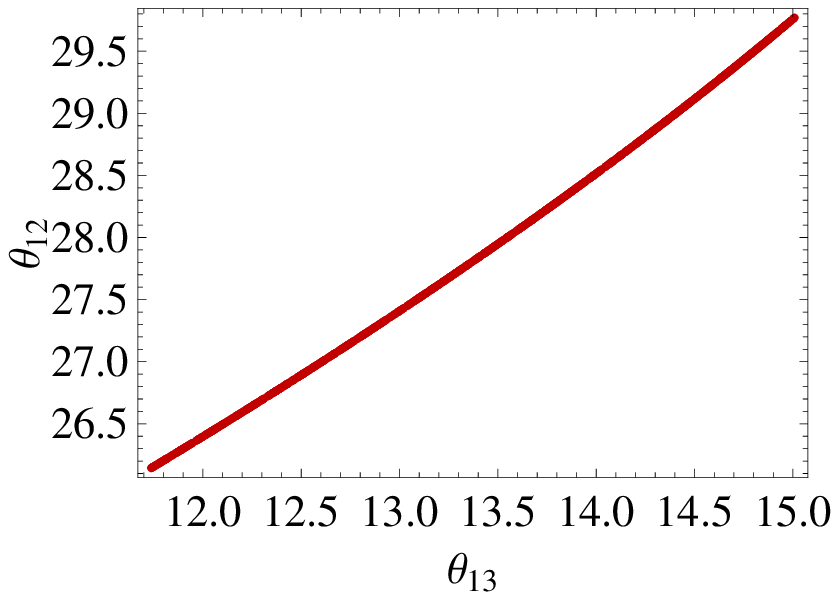}
		\includegraphics[width=7.3cm]{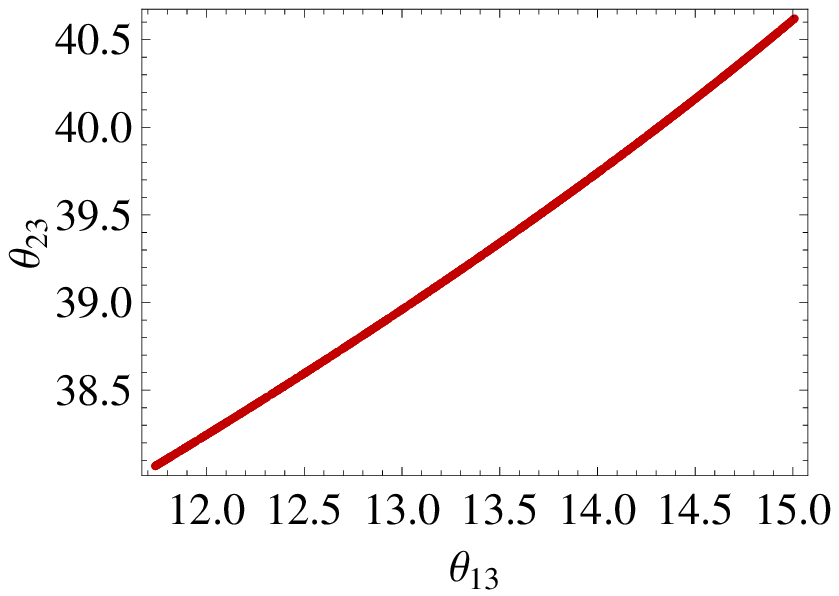}
		\includegraphics[width=7.3cm]{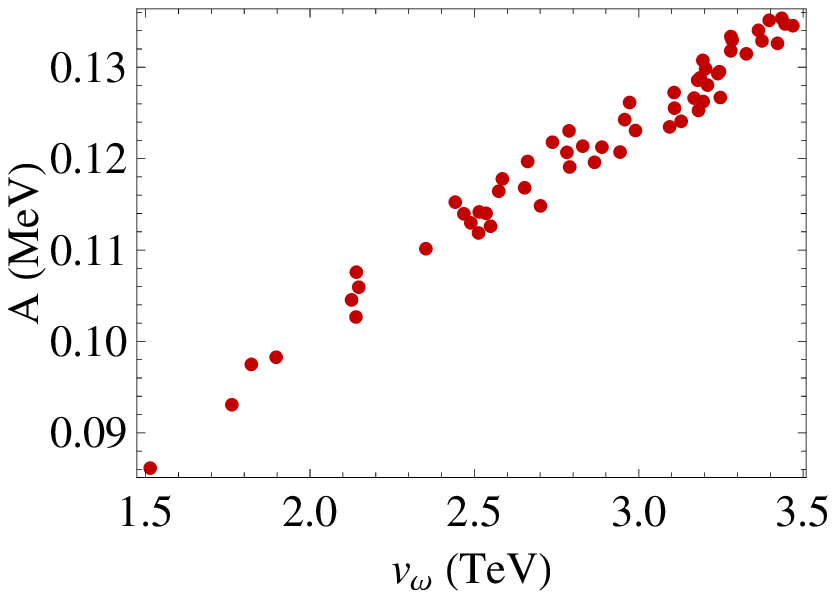}
		\includegraphics[width=7.3cm]{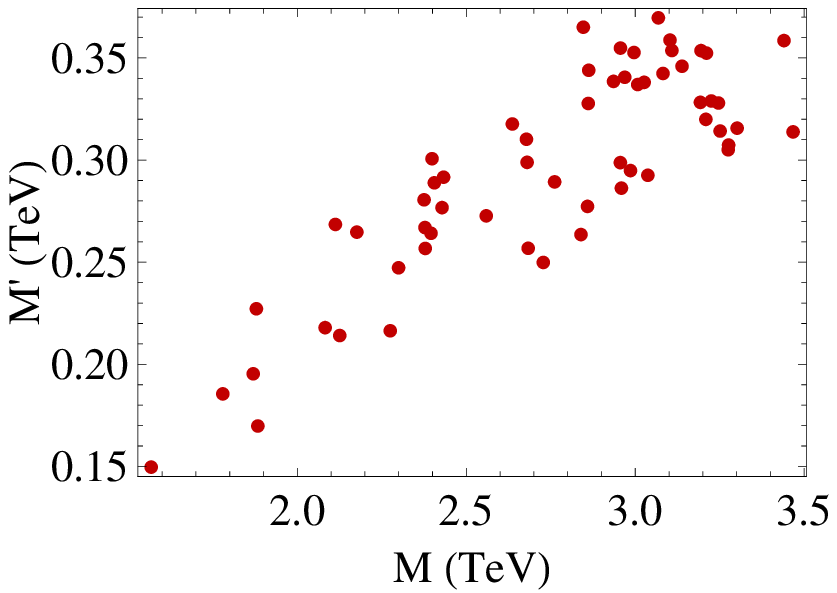}
		\includegraphics[width=7.3cm]{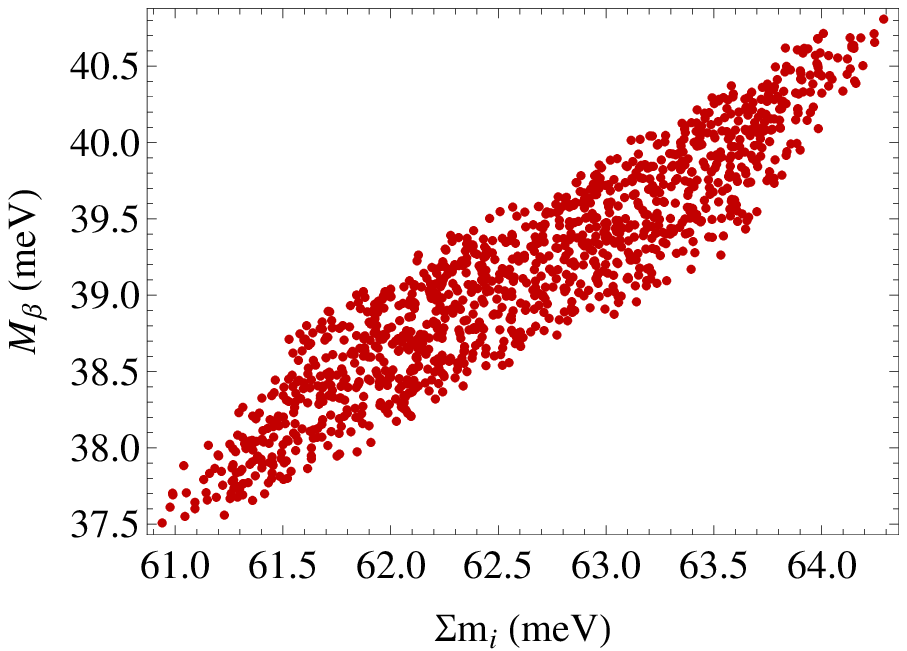}
		\includegraphics[width=7.3cm]{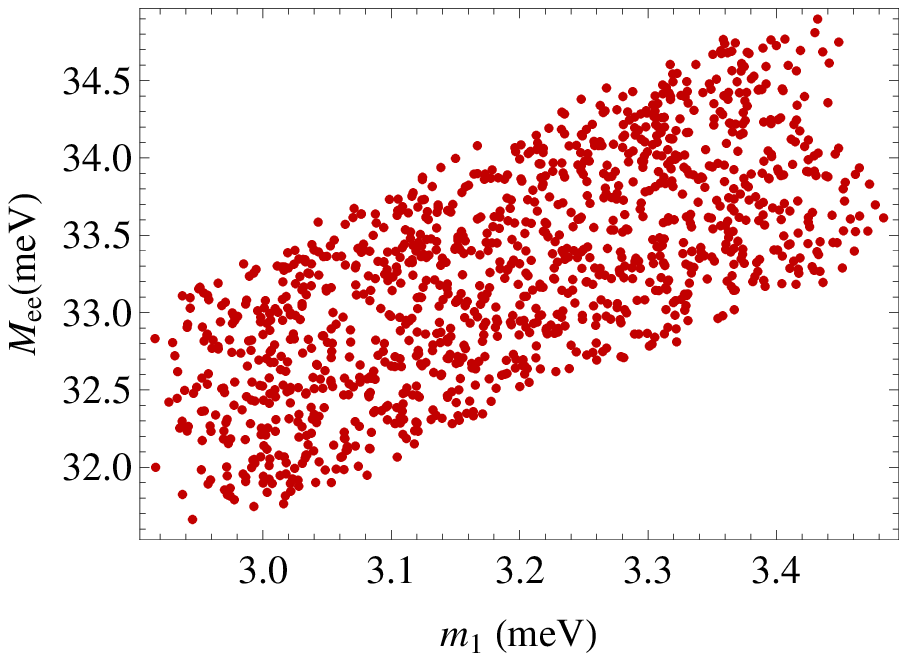}
	\caption{Scatter plots for $z=2.06$ with $s_{12l}\approx  -0.3,\; s_{13l}\approx  -0.001$, and  $s_{23l} \approx -0.061$. In the neutrino sector, we assume that $\tau=0.05 $. (meV $\equiv 10^{-3}$ eV)}
	\label{plot2}
\end{figure}

\begin{figure}[b]
	\centering
		\includegraphics[width=7.3cm]{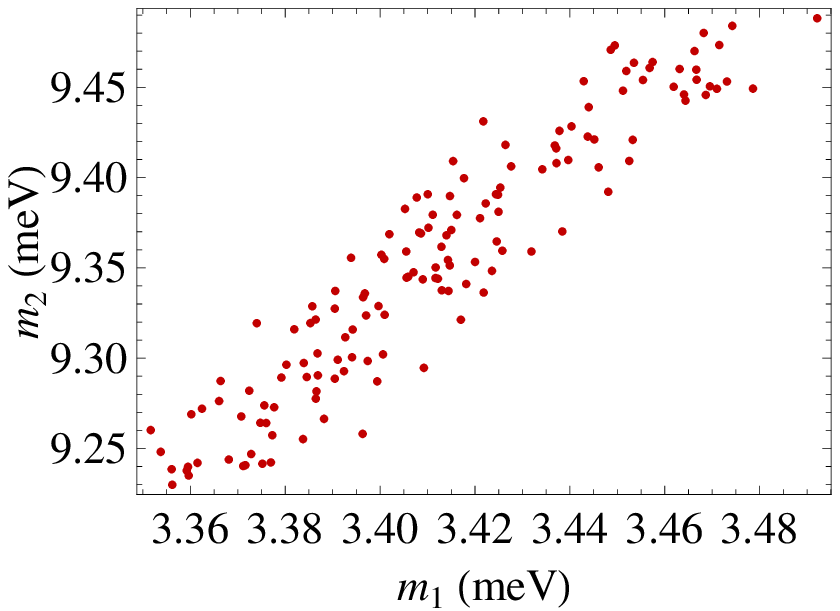}
		\includegraphics[width=7.3cm]{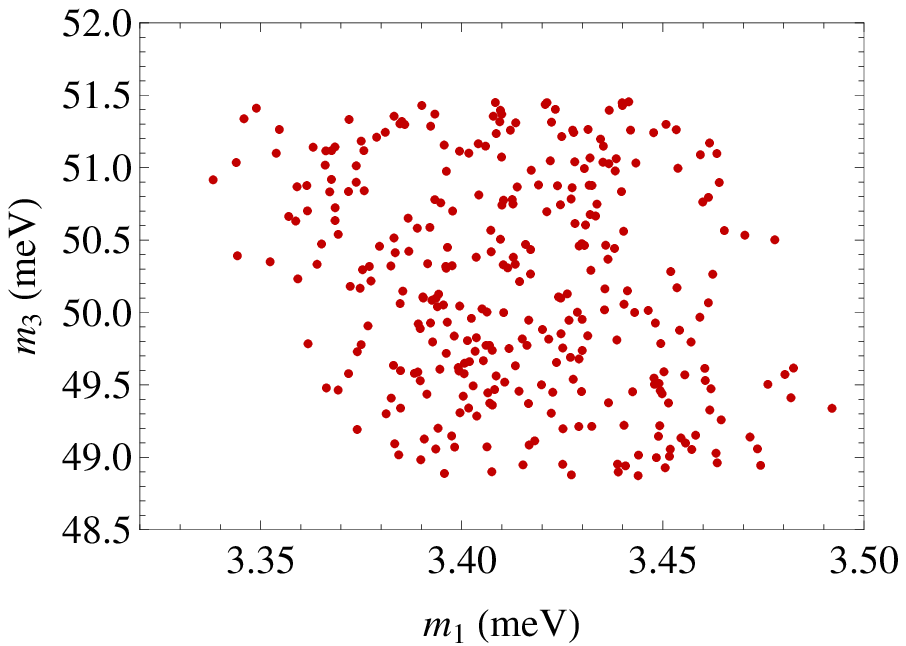}
		\includegraphics[width=7.3cm]{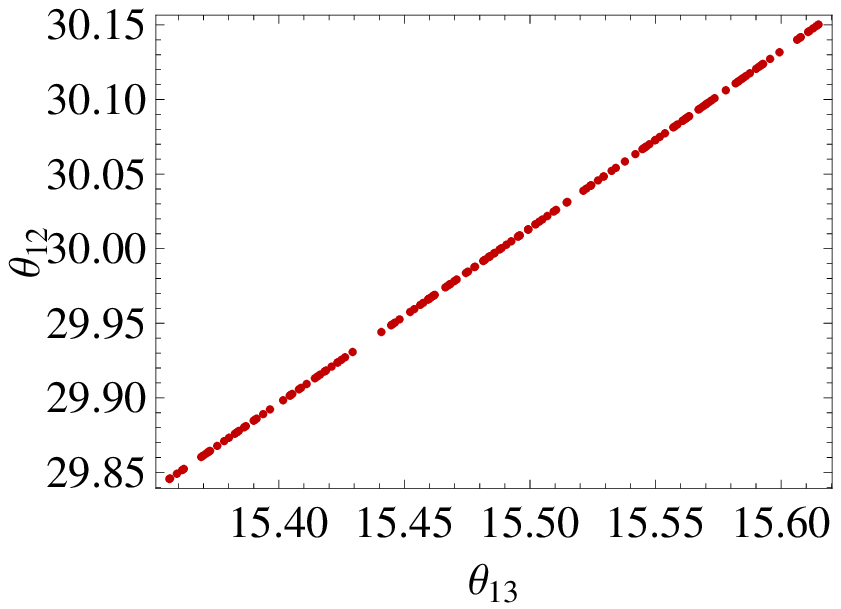}
		\includegraphics[width=7.3cm]{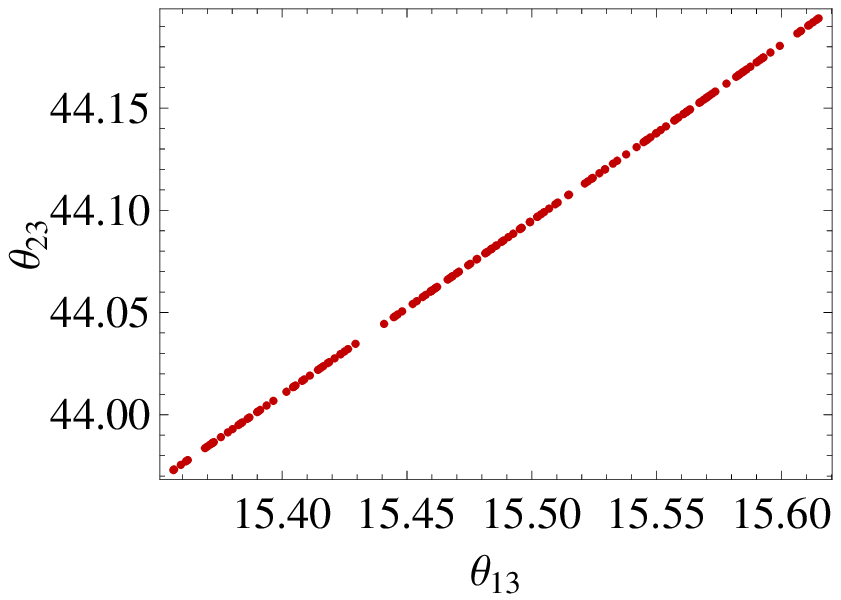}
		\includegraphics[width=7.3cm]{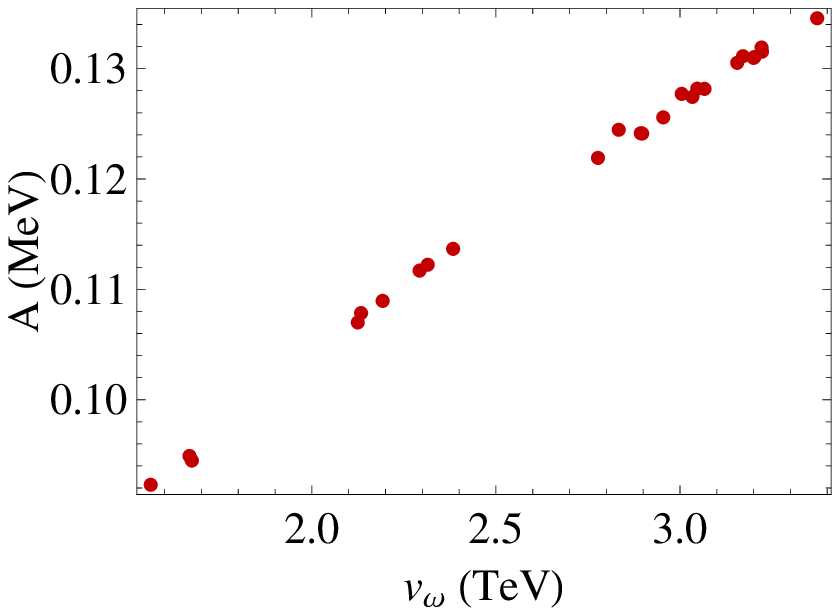}
		\includegraphics[width=7.3cm]{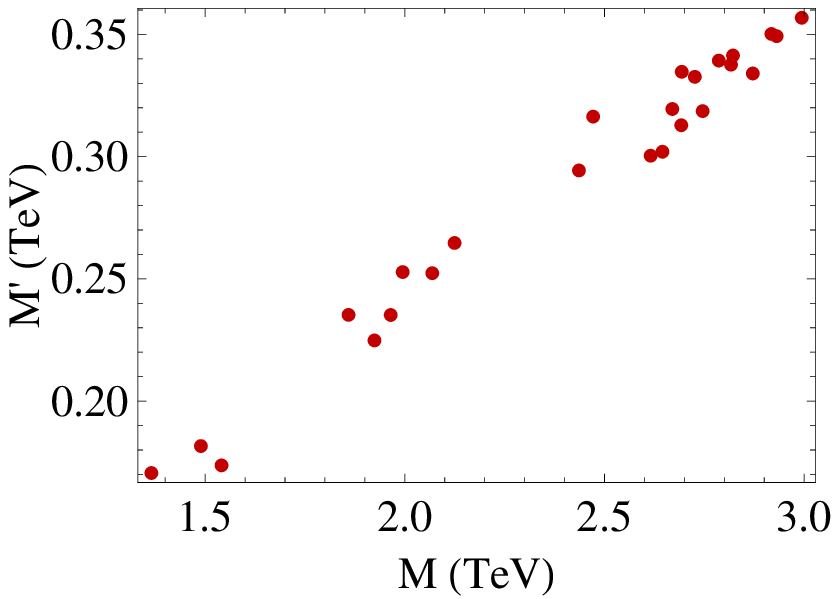}
		\includegraphics[width=7.3cm]{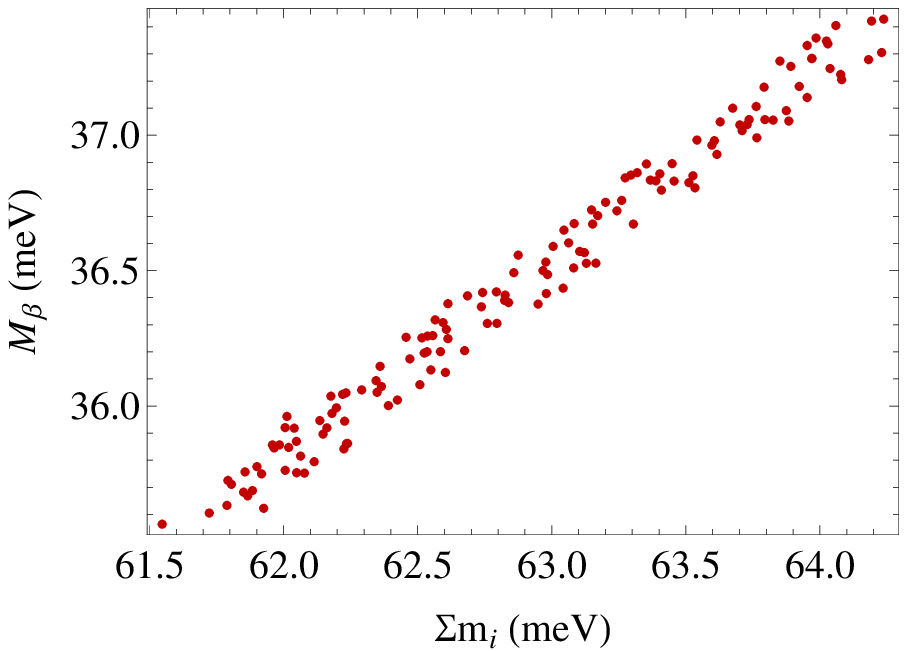}
		\includegraphics[width=7.3cm]{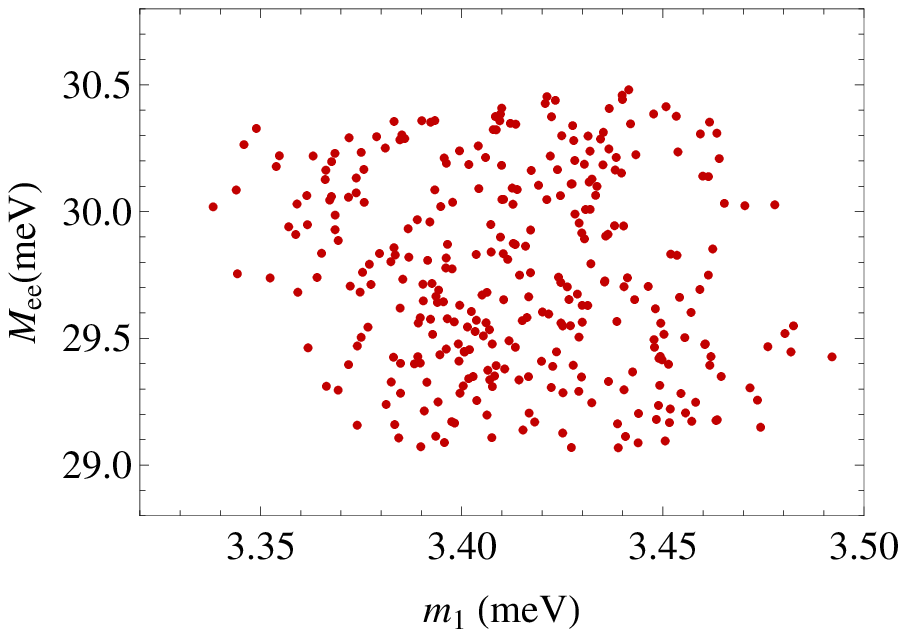}
	\caption{Scatter plots for $z=2.2$ with $s_{12l}\approx  -0.2,\; s_{13l}\approx  -0.00075$, and  $s_{23l} \approx -0.065$. In the neutrino sector, we assume that $\tau=0.1 $. (meV $\equiv 10^{-3}$ eV)}
	\label{plot3}
\end{figure}

In Figs.~(\ref{plot1}, \ref{plot2}, \ref{plot3}) we assume specific values of $z$ with the corresponding correction mixing angles $(s_{13l},\; s_{12l},\; s_{23l})$ and plot the absolute masses and the mixing angles which satisfy the neutrino mixing constraints. By choosing a value for the symmetry breaking term $\tau$, we plot the parameters $(v_w,\; A,\; M,\; M')$ that satisfy the squared mass difference measurements. This model supports the normal mass hierarchy as shown in the graphs with the scale of the neutrino masses in the few meV to $ \sim$ 50 meV range. The results agree with the recent T2K data which find a relatively large $\theta_{13}$. The graphs show that the see-saw scale $(M,\; M')$ are in the TeV range, and the second Higgs that couples to the right-handed neutrinos has v.e.v $v_2$, included in $A$, in the MeV scale. Also, they indicate that the v.e.v of the singlet scalar fields $v_w$ is in the TeV scale. The graphs show that $\Sigma m_i\approx 0.06$ eV and $ M_{ee} < M_\beta $ and $M_{ee} < 0.35$ eV \cite{Mohapatra:2006gs}. Various other mechanisms to generate the neutrino masses with TeV scale new physics are mentioned in Ref.~\cite{Chen:2011de}.

\section{Conclusion}


In this paper we presented a model for leptonic mixing which accommodates the sizable neutrino 
mixing angle $\theta_{13}$,  recently measured by the T2K and MINOS experiments. We worked in a basis where the charged lepton mass matrix is not diagonal and
 proposed an explicit structure for the charged lepton mass matrix which is 2-3 symmetric except for a single breaking of this symmetry by the muon mass.
 We identified a flavor symmetric limit for the mass matrices where the first generation is decoupled from the other two in the charged lepton sector while in the neutrino sector the third generation is decoupled from the first two
 generations. The leptonic mixing in the symmetric limit 
was shown to have, among other structures, the bi-maximal (BM) and the tri-bimaximal (TBM) mixing. 

 A model that extended the SM by three right handed neutrinos, an extra Higgs doublet, and two singlet scalars was introduced to generate the leptonic mixing. In the symmetric limit the model had two $Z_2$ symmetries in addition to the $\mu-\tau$ symmetry and the BM leptonic mixing was obtained when the two singlet scalars got equal v.e.v's.

Symmetry breaking effects were included in the charged lepton sector via higher dimensional operators that generated a $ \mu-\tau$ symmetric mass matrix 
except for a single breaking due to the finite muon mass. In the neutrino sector, symmetry breaking was included via  slightly different v.e.v's for the two singlet scalars. To explain the  $\Delta m^2$ data two different Majorana mass terms, one for $\nu_e$ and one for $ \nu_{\mu}$ and $\nu_{\tau}$, was used  keeping in mind that the $\mu-\tau$ symmetry fixes the Majorana mass terms for the $\nu_{\mu}$ and $\nu_{\tau}$ to be the same.

A fit to the experimental measurements showed that our model predicted normal hierarchy for the neutrino masses with the masses being in the few  meV to $ \sim $ 50 meV range. The Majorana mass terms as well as the v.e.v's of the singlet scalar fields were predicted to be in the TeV scale and consequently the v.e.v of the second Higgs doublet was shown to be in the MeV range. We calculated predictions for the mass-dependent observables $(\Sigma m_i)$, $(M_\beta)$ and $(M_{ee})$. We found that $\Sigma m_i\approx 0.06$ eV, $ M_{ee} < M_\beta $, and $M_{ee} < 0.35$ eV.

\section{Acknowledgements}
This work was
supported in part by the US-Egypt Joint Board on Scientific and Technological
Co-operation award (Project ID: 1855) administered by the US
Department of Agriculture, summer grant from the College of Liberal Arts, 
University of Mississippi and
 in part by the National Science Foundation under Grant No. 1068052 and 1066293 and the hospitality of the Aspen Center for Physics.


\newpage

\appendix
\begin{center}
 \section*{Appendices}
\end{center}

\section{Charged lepton sector\label{app:charged}}



%
We  analytically calculate the deviation of the leptonic mixing from the symmetric limit due to corrections 
from the charged lepton sector. We, here, are going to determine the sizes for the Yukawa matrix elements in $Y^L$ in Eq.~\ref{Yukawa-matrix-6}. We first consider the breaking of the 2-3 symmetry in the charged lepton sector via the introduction of a higher dimensional operator that generates the muon mass
\beq
O_1=c y_2\bar{D}_{\mu_L}\mu_R \phi_1 \frac{\phi_1 \phi_1^\dagger}{\Lambda^2}.
\label{app-1}
\eeq
Thus, we consider the Yukawa matrix,
\bea
 Y^L_{23} &= &\pmatrix{l_e & 0 &  0 \cr 0 & \frac{1}{2}{l_{T}}(1+2\kappa_l) & \frac{1}{2}{l_{T} }\cr 0
&\frac{1}{2}{l_{T}} & \frac{1}{2}{l_{T}}}.\
 \label{23dsymbreak1}
 \eea
The structure above breaks the $2-3$ symmetry because of the correction to the $22$ element. Note that we do not break  the $2-3$ symmetry  in the  $23$ element  so that  the Yukawa matrix remains  symmetric. The matrix $Y^L_{23}$  is now diagonalized by the unitary matrix, $U_l= W^l_{23} R^l_{23}$. 
%
Applying the relation $(Y^{L}_{23})_{23}=(Y^{L}_{23})_{33}$ in Eq.~\ref{23dsymbreak1} to $Y^L_{23}  =  U_l Y^L_{diag} U_l^{\dagger}$ leads to
 \bea
 t_{23l} & = & \frac{1}{2} \left[ z_\mu-1 + \sqrt{z_\mu^2-6z_\mu+1} \right], \
 \label{theta23solnd}
 \eea
 where $t_{23l}\equiv\tan{\theta_{23l}}$ and we have chosen the solution that leads to small angle  $\theta_{23l}$  and to small flavor symmetry breaking.  Keeping terms up to first order in $z_\mu$ we get
 \bea
t_{23l} & \approx  & -z_\mu. \
 \label{theta23solnapproxd}
 \eea
  We further obtain for $\kappa_l$ and $l_T$ in Eq.~\ref{23dsymbreak1},
 \bea
 \kappa_l &=& -\tan{2 \theta_{23l}} \approx 2 z_\mu, \nonumber\\
 l_{T} & = & (l_{\tau}-l_{\mu})\cos{ 2 \theta_{23l}}.\
 \label{chi}
 \eea
Comparing the above equation with Eq.~\ref{app-1}, the size of the higher dimensional operator can be estimated as
\beq
\frac{c v_1^2}{2\Lambda^2} \approx 2 z_\mu.
\eeq
Since $v_1\approx 250$ GeV therefore the scale of $\Lambda $ is in the TeV range.

To obtain a realistic charged lepton matrix, we take into account the mixing involving the second and the third generations in the full Yukawa matrix 
\beq
O_2= y' \left( \bar{D}_{e_L}\mu_R - \bar{D}_{e_L}\tau_R + \bar{D}_{\mu_L}e_R - \bar{D}_{\tau_L}e_R\right) \phi_1 \frac{\phi_1 \phi_1^\dagger}{\Lambda^2}. 
\label{app-2}
\eeq
Thus, the full mixing matrix will be give by
\bea
 Y^L &= &\pmatrix{l_{11} & l_{12} &  -l_{12} \cr l_{12} & \frac{1}{2}{l_{T}}(1+2\kappa_l) & \frac{1}{2}{l_{T} }\cr -l_{12}
&\frac{1}{2}{l_{T}} & \frac{1}{2}{l_{T}}}.\
\label{Yukawa-matrix-8}
\eea
 We will assume that the Yukawa matrix $Y^{L}$ is now diagonalized by the unitary matrix $U_l$ given by 
 \bea
 U_l & = & W^l_{23}R^{l}_{23}R^{l}_{13}R^{l}_{12}. \
 \label{diagmat}
 \eea
From the Yukawa matrix (\ref{Yukawa-matrix-8}), one can find  the two relations
\bea
Y_{12}&=&-Y_{13}, \nonumber\\
Y_{22} & = & \frac{1}{2}(Y_{23}+Y_{33})(1+2 \kappa_l).
\eea
Applying the above two relations to 
\beq
Y^L  =  U_l Y^L_{diag} U_l^{\dagger}
\label{YL-final}
\eeq
using Eq.~\ref{diagmat}, one can obtain the solutions
\bea
s_{12l} & \approx & \pm c_{23l} \sqrt{\frac{z_{\mu}-2 \kappa_{l}+(-3+3 z_{\mu}-2 \kappa_l) c_{23l} s_{23l}+2z_{\mu}\kappa_l c_{23l} s_{23l}}{z_{\mu}-z_{\mu}^2 (3+2 \kappa_l) c_{23l} s_{23l}}},   \nonumber \\
s_{13l} & \approx & \pm \sqrt{z_{\mu}} c_{12l} s_{23l} \sqrt{\frac{z_{\mu}-2\kappa_l +(-3+3 z_{\mu}-2\kappa_l)c_{23l} s_{23l}+2z_{\mu}\kappa_l c_{23l} s_{23l}}{1-z_{\mu}(3+2 \kappa_l)  c_{23l} s_{23l}}}.
\label{s12l-s13l}
\eea
 By comparing Eqs.~(\ref{Yukawa-matrix-8}, \ref{YL-final}), one can get the matrix element $l_{12}$ after substituting Eqs.~(\ref{theta23solnapproxd}, \ref{chi}, \ref{s12l-s13l}) up to the first order in $z_\mu$ 
\beq
l_{12} \approx \sqrt{\frac{z_\mu}{2}} (l_e - l_\mu).
\eeq
The leptonic mixing matrix is now given by
\bea
U_{PMNS}&= U^{\dagger}_\ell U_\nu, \
\label{nuleptonfinal}
\eea
where
$U_\ell  =  W^l_{23} R^l_{23} R^l_{13} R^l_{12}$ and $U_{\nu}=W^{\nu}_{12}$.


\begin{thebibliography}{99}

\expandafter\ifx\csname url\endcsname\relax
  \def\url#1{\texttt{#1}}\fi
\expandafter\ifx\csname urlprefix\endcsname\relax\def\urlprefix{URL }\fi
\providecommand{\eprint}[2][]{\url{#2}}

\bibitem{Bahcall:2004cc-1}
  J.~N.~Bahcall,
  arXiv:physics/0406040;
  R.~N.~Mohapatra {\it et al.},
  arXiv:hep-ph/0510213;
  R.~N.~Mohapatra and A.~Y.~Smirnov,
  Ann.\ Rev.\ Nucl.\ Part.\ Sci.\  {\bf 56} (2006) 569
  [arXiv:hep-ph/0603118];
  S.~F.~King,
  Rept.\ Prog.\ Phys.\  {\bf 67} (2004) 107
  [arXiv:hep-ph/0310204];
  G.~Altarelli and F.~Feruglio,
  New J.\ Phys.\  {\bf 6} (2004) 106
  [arXiv:hep-ph/0405048].


\bibitem{PMNS-1}
Z. Maki, M. Nakagawa and S. Sakata, Prog. Theo. Phys. {\bf 28} (1962)
247;
B.~W.~Lee, S.~Pakvasa, R.~E.~Shrock and H.~Sugawara,
Phys.\ Rev.\ Lett.\  {\bf 38} (1977) 937
[Erratum-ibid.\  {\bf 38} (1977) 1230].
W.-M.~Yao {\it et al.}\ [Particle Data Group Collaboration],



\bibitem{BM-1}
 V.~Barger, S.~Pakvasa, T.~Weiler and K.~Whisnant, Phys. Lett. B \textbf{437}, 107 (1998);  N.~Li and B.~Q.~Ma,
  Phys.\ Lett.\  B {\bf 600}, 248 (2004)
  [arXiv:hep-ph/0408235].
  I.~Stancu and D.~V.~Ahluwalia,
  Phys.\ Lett.\  B {\bf 460}, 431 (1999)
  [arXiv:hep-ph/9903408].


\bibitem{TBM-1}
P.~F.~Harrison, D.~H.~Perkins, and W.~G.~Scott, Phys.~Lett.~B530 (2002) 167, 
\eprint{hep-ph/0202074}; Phys.~Lett.~B458 (1999) 79, \eprint{hep-ph/9904297};
Z.~Xing Phys.~Lett.~B533 (2002) 85, \eprint{hep-ph/0204049};
X.\ He and A.\ Zee, Phys.\ Lett.\ B560 (2003) 87, \eprint{hep-ph/0301092};
E.\ Ma, Phys.\ Rev.\ Lett.\ 90 (2003) 221802, \eprint{hep-ph/0303126};
Phys.\ Lett.\ B583 (2004) 157, \eprint{hep-ph/0308282};
Phys.\ Rev.\ D70 (2004) 031901, \eprint{hep-ph/0404199};
C.\ I.\ Low and R.\ R.\ Volkas, Phys.\ Rev. D68 (2003) 033007, \eprint{hep-ph/0305243};
G.\ Altarelli and F.\ Feruglio, Nucl. Phys.\ B 720 (2005) 64. \eprint{hep-ph/0504165}.


\bibitem{Abe:2011sj}
  K.~Abe {\it et al.} [ T2K Collaboration ],  [arXiv:1106.2822 [hep-ex]].
  


\bibitem{MINOS}
L.~Whitehead [MINOS Collaboration], ``\textit{Recent results from MINOS}'', Joint Experimental-Theoretical Seminar (24 June 2011, Fermilab, USA). Websites: theory.fnal.gov/jetp, http://www-numi.fnal.gov/pr-plots/


\bibitem{MINOS-formal}
\textit{P.~Adamson et al.} [MINOS Collaboration], ``\textit{Improved search for muon-neutrino to electron-neutrino oscillations in MINOS},'' arXiv:1108.0015 [hep-ex].



\bibitem{T2K-cited-CL-NonDiag-1}

 D.~Marzocca, S.~T.~Petcov, A.~Romanino, M.~Spinrath,
    [arXiv:1108.0614 [hep-ph]].
  Y.~H.~Ahn, H.~-Y.~Cheng, S.~Oh,
    [arXiv:1107.4549 [hep-ph]].
  R.~d.~A.~Toorop, F.~Feruglio, C.~Hagedorn,
    [arXiv:1107.3486 [hep-ph]].
     P.~S.~Bhupal Dev, R.~N.~Mohapatra, M.~Severson,
    [arXiv:1107.2378 [hep-ph]].
      S.~Dev, S.~Gupta, R.~R.~Gautam,
    [arXiv:1107.1125 [hep-ph]].
    D.~Meloni,
    [arXiv:1107.0221 [hep-ph]].
  N.~Haba, R.~Takahashi,
  Phys.\ Lett.\  {\bf B702}, 388-393 (2011).
  [arXiv:1106.5926 [hep-ph]].
  S.~Zhou,
    [arXiv:1106.4808 [hep-ph]].
      X.~-G.~He, A.~Zee,
    [arXiv:1106.4359 [hep-ph]].
      Y.~-j.~Zheng, B.~-Q.~Ma,
    [arXiv:1106.4040 [hep-ph]].
      H.~Ishimori, T.~Kobayashi,
    [arXiv:1106.3604 [hep-ph]].
      M.~-C.~Chen, K.~T.~Mahanthappa, A.~Meroni, S.~T.~Petcov,
    [arXiv:1109.0731 [hep-ph]].
    



  
\bibitem{T2K-cited-CL-Diag-2}
 S.~Kumar,
    [arXiv:1108.2137 [hep-ph]].
  H.~Fritzsch, Z.~-z.~Xing, S.~Zhou,
    [arXiv:1108.4534 [hep-ph]].
  S.~Antusch, S.~F.~King, C.~Luhn, M.~Spinrath,
    [arXiv:1108.4278 [hep-ph]].
  M.~Huang, D.~Liu, J.~-C.~Peng, S.~D.~Reitzner, W.~-C.~Tsai,
  [arXiv:1108.3906 [hep-ph]].
  T.~Araki, C.~-Q.~Geng,
  [arXiv:1108.3175 [hep-ph]].
  Riazuddin,
    [arXiv:1108.1469 [hep-ph]].
  T.~Schwetz, M.~Tortola, J.~W.~F.~Valle,
    [arXiv:1108.1376 [hep-ph]].
  S.~-F.~Ge, D.~A.~Dicus, W.~W.~Repko,
  [arXiv:1108.0964 [hep-ph]].
  G.~J.~Mathews, T.~Kajino, W.~Aoki, W.~Fujiya,
    [arXiv:1108.0725 [astro-ph.HE]].
  S.~F.~King, C.~Luhn,
    [arXiv:1107.5332 [hep-ph]].
 M.~-C.~Chen, K.~T.~Mahanthappa,
    [arXiv:1107.3856 [hep-ph]].
  W.~Rodejohann, H.~Zhang, S.~Zhou,
    [arXiv:1107.3970 [hep-ph]].
  S.~Antusch, V.~Maurer,
    [arXiv:1107.3728 [hep-ph]].
  X.~Chu, M.~Dhen, T.~Hambye,
    [arXiv:1107.1589 [hep-ph]].
  H.~Zhang, S.~Zhou,
    [arXiv:1107.1097 [hep-ph]].
  W.~Chao, Y.~-j.~Zheng,
    [arXiv:1107.0738 [hep-ph]].
  S.~Morisi, K.~M.~Patel, E.~Peinado,
    [arXiv:1107.0696 [hep-ph]].
  G.~L.~Fogli, E.~Lisi, A.~Marrone, A.~Palazzo, A.~M.~Rotunno,
    [arXiv:1106.6028 [hep-ph]].
  T.~Araki,
  Phys.\ Rev.\  {\bf D84}, 037301 (2011).
  [arXiv:1106.5211 [hep-ph]].
 A.~B.~Balantekin,
    [arXiv:1106.5021 [hep-ph]].
  S.~F.~King,
    [arXiv:1106.4239 [hep-ph]].
 E.~Ma, D.~Wegman,
  Phys.\ Rev.\ Lett.\  {\bf 107}, 061803 (2011).
  [arXiv:1106.4269 [hep-ph]].
  J.~-M.~Chen, B.~Wang, X.~-Q.~Li,
    [arXiv:1106.3133 [hep-ph]].
  Z.~-z.~Xing,
    [arXiv:1106.3244 [hep-ph]].
  N.~Qin, B.~Q.~Ma,
  Phys.\ Lett.\  {\bf B702}, 143-149 (2011).
  [arXiv:1106.3284 [hep-ph]].
  J.~Barry, W.~Rodejohann, H.~Zhang,
  JHEP {\bf 1107}, 091 (2011).
  [arXiv:1105.3911 [hep-ph]].
   Q.~-H.~Cao, S.~Khalil, E.~Ma, H.~Okada,
  [arXiv:1108.0570 [hep-ph]].



\bibitem{He:2011kn} 
  H.~-J.~He and F.~-R.~Yin,
  Phys.\ Rev.\ D {\bf 84}, 033009 (2011)
  [arXiv:1104.2654 [hep-ph]].



\bibitem{CCC-1}
  P.~H.~Frampton, S.~T.~Petcov and W.~Rodejohann,
  Nucl.\ Phys.\ B {\bf 687} (2004) 31, hep-ph/0401206;
  G.~Altarelli, F.~Feruglio and I.~Masina,
  Nucl.\ Phys.\ B {\bf 689} (2004) 157,
  hep-ph/0402155;
  S.~Antusch and S.~F.~King,
  Phys.\ Lett.\ B {\bf 591} (2004) 104,
  hep-ph/0403053;
 F.~Feruglio,
  Nucl.\ Phys.\ Proc.\ Suppl.\  {\bf 143} (2005) 184
  [Nucl.\ Phys.\ Proc.\ Suppl.\  {\bf 145} (2005) 225],
  hep-ph/0410131;
  R.~N.~Mohapatra and W.~Rodejohann,
 hep-ph/0507312.
  S.~Antusch and S.~F.~King,
  Phys.\ Lett.\  B {\bf 659} (2008) 640
  [arXiv:0709.0666 [hep-ph]].

\bibitem{datta-1}
  A.~Datta,
  Phys.\ Rev.\  D {\bf 78}, 095004 (2008)
  [arXiv:0807.0795 [hep-ph]];
A.~Datta,
  Phys.\ Rev.\  {\bf D74}, 014022 (2006).
  [hep-ph/0605039];
A.~Datta, P.~J.~O'Donnell,
  Phys.\ Rev.\  {\bf D72}, 113002 (2005).
  [hep-ph/0508314];
A.~S.~Joshipura, B.~P.~Kodrani,
  Phys.\ Rev.\  {\bf D82}, 115013 (2010).
  [arXiv:1004.3637 [hep-ph]].


\bibitem{datta1-1} 
S.~Baek, A.~Datta, P.~Hamel, O.~F.~Hernandez and D.~London,
  Phys.\ Rev.\  D {\bf 72}, 094008 (2005)
  [arXiv:hep-ph/0508149].
S.~Baek, P.~Hamel, D.~London, A.~Datta and D.~A.~Suprun,
  Phys.\ Rev.\  D {\bf 71}, 057502 (2005)
  [arXiv:hep-ph/0412086];
 A.~Datta, M.~Imbeault, D.~London, V.~Page, N.~Sinha, R.~Sinha,
  Phys.\ Rev.\  {\bf D71}, 096002 (2005).
  [hep-ph/0406192];
 A.~Datta, D.~London,
  Phys.\ Lett.\  {\bf B595}, 453-460 (2004).
  [hep-ph/0404130].
   A.~Datta,
  Phys.\ Rev.\  D {\bf 66}, 071702 (2002)
  [arXiv:hep-ph/0208016];
A.~Datta, X.~G.~He and S.~Pakvasa,
  Phys.\ Lett.\  B {\bf 419}, 369 (1998)
  [arXiv:hep-ph/9707259].


\bibitem{TBM-5}
C.\ I.\ Low and R.\ R.\ Volkas, Phys.\ Rev. D68 (2003) 033007, \eprint{hep-ph/0305243};









\bibitem{2HDM-1}
 S.~Andreas, O.~Lebedev, S.~Ramos-Sanchez {\it et al.},
  JHEP {\bf 1008}, 003 (2010).
  [arXiv:1005.3978 [hep-ph]].
A.~Rashed, M.~Duraisamy, A.~Datta,
  Phys.\ Rev.\  {\bf D82}, 054031 (2010).
  [arXiv:1004.5419 [hep-ph]]. 
R.~Dermisek, J.~F.~Gunion,
  Phys.\ Rev.\  {\bf D81}, 055001 (2010).
  [arXiv:0911.2460 [hep-ph]].



\bibitem{Grimus:2003kq-1}
  W.~Grimus, L.~Lavoura,
  Phys.\ Lett.\  {\bf B572}, 189-195 (2003).
  [hep-ph/0305046].
   W.~Grimus, L.~Lavoura,
    J.\ Phys.\ G {\bf G30}, 73-82 (2004).
    [hep-ph/0309050].
    W.~Grimus, A.~S.~Joshipura, S.~Kaneko, L.~Lavoura, M.~Tanimoto,
      JHEP {\bf 0407}, 078 (2004).
      [hep-ph/0407112].
        W.~Grimus, S.~Kaneko, L.~Lavoura, H.~Sawanaka, M.~Tanimoto,
        JHEP {\bf 0601}, 110 (2006).
        [hep-ph/0510326].




\bibitem{mutau-1} 
T. Fukuyama and H. Nishiura,
hep-ph/9702253; in Proceedings of the International 
Workshop on Masses and 
Mixings of Quarks and Leptons,  Shizuoka, Japan, 1997, 
edited by Y.~Koide (World Scientific, Singapore, 1998), p. 252;
R.N. Mohapatra and S. Nussinov, Phys. Rev. {\bf D60}, 013002 (1999);
E. Ma and  M. Raidal, Phys. Rev. Lett. {\bf 87}, 011802 (2001);
C. S. Lam, hep-ph/0104116; 
T. Kitabayashi and
M. Yasue, Phys.Rev. {\bf D67} 015006 (2003);
 Y. Koide,  Phys.Rev. {\bf D69}, 093001 (2004);
R. N. Mohapatra, SLAC Summer
Inst. lecture; http://www-conf.slac.stanford.edu/ssi/2004; hep-ph/0408187;
JHEP, {\bf 0410}, 027 (2004);
  W. Grimus, A. S.Joshipura, S. Kaneko, L.
Lavoura, H. Sawanaka, M. Tanimoto, hep-ph/0408123; 
A.~Ghosal,
  Mod.\ Phys.\ Lett.\ A {\bf 19}, 2579 (2004).




\bibitem{seesaw-1}
M. Gell-Mann, P. Ramond, and R. Slansky,
in \textit{Supergravity, Proceedings of the Workshop, Stony Brook,
New York, 1979},
eds. F. van Nieuwenhuizen and D. Freedman
(North Holland, Amsterdam, 1979);
T. Yanagida,
in \textit{Proceedings of the Workshop
on Unified Theories and Baryon Number in the Universe},
Tsukuba, Japan, 1979,
eds. O. Sawada and A. Sugamoto
(KEK report no. 79--18, Tsukuba, 1979); \}
R.N. Mohapatra and G. Senjanovi\'c, Phys. Rev. Lett. 44 (1980) 912.



\bibitem{Barry:2010yk}
  J.~Barry and W.~Rodejohann,
  Nucl.\ Phys.\  B {\bf 842}, 33 (2011)
  [arXiv:1007.5217 [hep-ph]].





\bibitem{Schwetz:2008er}
  T.~Schwetz, M.~A.~Tortola and J.~W.~F.~Valle,
  New J.\ Phys.\  {\bf 10}, 113011 (2010)
  [arXiv:0808.2016v3 [hep-ph]].




\bibitem{king}
S.~F.~King,
  Phys.\ Lett.\  B {\bf 659}, 244 (2008)
  [arXiv:0710.0530 [hep-ph]].




\bibitem{Mohapatra:2006gs}
  R.~N.~Mohapatra and A.~Y.~Smirnov,
  Ann.\ Rev.\ Nucl.\ Part.\ Sci.\  {\bf 56}, 569 (2006)
  [arXiv:hep-ph/0603118].

 




\bibitem{Chen:2011de}
  M.~-C.~Chen, J.~Huang, [arXiv:1105.3188 [hep-ph]].
  
  





\end{thebibliography}
\end{document}